\documentclass{aa}  

\usepackage{graphicx}
\usepackage{txfonts}
\usepackage{lipsum}
\usepackage{subcaption}         % necessary for continued figures, example in section 3
\usepackage{lscape}             % to rotate a single page table, example in appendix.
\usepackage{placeins}           % useful with \FloatBarrier, to keep 
\usepackage{CJKutf8}
\usepackage{xcolor}

\newcommand{\cjk}[1]{{\begin{CJK*}{UTF8}{bsmi}#1\end{CJK*}}}

\begin{document}
%\begin{CJK*}{UTF8}{bsmi}

%%%%%%%%%%%%%%%%%%%%%%%%%%%%%%%%%%%%%%%%
% if you use custom commands in your title,
% ensure to check your title when submitting!
%%%%%%%%%%%%%%%%%%%%%%%%%%%%%%%%%%%%%%%%
   \title{Why Do Stars Turn Red?}

   \subtitle{I. Post-Main-Sequence Expansion Mechanism}

%%%%%%%%%%%%%%%%%%%%%%%%%%%%%%%%%%%%%%%%
% Please separate each author with the \and command
%
% Please do not include ORCIDs next to author names.
% Only ORCIDs authenticated by individual authors in EDPS
% editorial system will be taken into account.
% ORCIDs included here will be removed.
%%%%%%%%%%%%%%%%%%%%%%%%%%%%%%%%%%%%%%%%

   \author{Po-Sheng Ou (\cjk{歐柏昇})\inst{1,2}\fnmsep\thanks{Corresponding author: psou@asiaa.sinica.edu.tw}
        \and Ke-Jung Chen (\cjk{陳科榮})\inst{1,3}
        }

   \institute{Institute of Astronomy and Astrophysics, Academia Sinica, No.1, Sec. 4, Roosevelt Rd., Taipei 106319, Taiwan, R.O.C.
   \and Department of Physics, National Taiwan University, No.1, Sec. 4, Roosevelt Rd.,  Taipei 106319, Taiwan, R.O.C.
   \and Heidelberg Institute for Theoretical Studies, Schloss-Wolfsbrunnenweg 35, Heidelberg 69118,
Germany}

   %\date{Received September 30, 20XX}

% \abstract{}{}{}{}{}
% 5 {} token are mandatory
 
  \abstract
   {In this series of papers, we address the long-standing question of why post-main-sequence stars expand into red giants (RGs) or red supergiants (RSGs).}
   {This paper aims to identify the key physical mechanism that drives stellar evolution toward the RG/RSG phase.} 
   {Using the Modules for Experiments in Stellar Astrophysics (MESA), we perform controlled numerical experiments by systematically varying stellar parameters in evolutionary models, and compare those that successfully evolve into RG/RSGs and those that do not.}
   {We show that envelope expansion toward the RG/RSG phase cannot be explained by energy absorption. Instead, it is governed by a refined form of the “mirror principle,” in which the stellar envelope responds oppositely to its inner boundary, defined by the outer edge of the hydrogen-burning shell, rather than directly to the helium core. This behavior arises naturally from hydrostatic equilibrium, as the burning shell establishes a moving, nearly constant-pressure inner boundary for the envelope. We identify two evolutionary pathways toward the RG/RSG phase that both follow this refined mirror principle: (1) direct envelope expansion during helium-core contraction, and (2) continued expansion after contraction ceases, driven by a decline in nuclear energy generation rate. The final approach to the RG/RSG phase is marked by a structural transition in the envelope, characterized by mass redistribution and the development of an extended convective region.}
   {We present a unified physical framework for envelope expansion toward the RG/RSG phase, based on the refined mirror principle and the final structural transition, and outline an evolutionary roadmap leading to the RG/RSG phase.}
   \keywords{Supergiants --
                Stars: interiors --
                Stars: evolution
               }

   \maketitle
   \nolinenumbers

\section{Introduction} \label{sec:intro}

When a star exhausts hydrogen in its core and leaves the main sequence, its helium (He) core contracts while its envelope expands. This expansion gives rise to red giants (RGs; initial masses $\lesssim 8\,M_{\odot}$) or red supergiants (RSGs; initial masses $\gtrsim 8\,M_{\odot}$), characterized by effective temperatures of $\sim 4,000\,$K. The inverse relationship between core contraction and envelope expansion is commonly referred to as the “mirror effect” or “mirror principle” \citep{Maeder2009,Kippenhahn2013}. Despite its long use, the physical mechanism that drives a stellar envelope to expand to the characteristic red state remains poorly understood.

Several theories have been proposed to explain why stars evolve into RGs/RSGs. Suggested drivers include the degree of core contraction \citep{Eggleton1991,Faulkner2005}, thermal imbalance set by core luminosity and envelope opacity \citep{Renzini1984,Renzini1992,Renzini1994,Ritossa1996,Renzini2023}, gradients in opacity and mean molecular weight \citep{Sugimoto2000}, and pressure-density contrasts between the shell and core \citep{MillerBertolami2022}. Other studies argue that no single factor is dominant \citep{Whitworth1989,Iben1993}. Despite these efforts, there remains no consensus on what ultimately drives stars to become red.

A useful clue comes from the concept of a “critical metallicity” for RSG formation, suggested by recent massive-star models \citep{Ou2023}. These studies show that not all stars expand into RSGs during core-He burning: only those above a metallicity threshold make the transition, while those below it remain as blue supergiants (BSGs) with smaller radii. In this work, we investigate the mechanism underlying this bifurcation in envelope expansion, determining why some stars evolve into BSGs while others become RSGs.

Owing to the complexity of the physics governing stellar expansion and the transition between blue and red evolutionary states, this study is presented in two parts. Paper~I (the present work) identifies the mechanism that drives the evolution of stars into RGs/RSGs using stellar evolution models, beginning with the establishment of a criterion that determines whether a star becomes red or remains blue. Paper~II \citep{paper2} provides the physical basis of these mechanisms by constructing steady-state envelope models.

The paper is organized as follows. Section~\ref{sec:methods} outlines the construction of our stellar models. In Section~\ref{sec:criterion}, we use $25\,M_{\odot}$ models to identify the criterion and mechanism driving evolution into the RSG stage. Section~\ref{sec:mass} extends this analysis to stars of different masses and metallicities. Section~\ref{sec:mirror} examines the physics underlying the mirror principle in envelope expansion, while Section~\ref{sec:envelope} explores roles of envelope properties—such as opacity and structural transitions—in shaping the envelope expansion. Finally, Section~\ref{sec:conclusion} summarizes our findings and highlights open questions for future work.

\section{Methods} \label{sec:methods}

To investigate the physical origin of RGs and RSGs, we compute stellar models with the Modules for Experiments in Stellar Astrophysics \citep[MESA;][]{Paxton2011,Paxton2013,Paxton2015,Paxton2018,Paxton2019,Jermyn2023}, version~10108. MESA is a one-dimensional stellar evolution code that solves the equations of stellar structure and composition, incorporating modules for nuclear reaction networks, equations of state, opacities, and convection via mixing-length theory (MLT). Its flexibility and extensive physics library make it particularly well suited for studying stellar evolution across a broad range of masses and metallicities.

We construct a grid of non-rotating stellar models with initial masses from $1$ to $30\,M_{\odot}$ in $1\,M_{\odot}$ intervals and 38 metallicities: $Z = (1,2,\dots,9)\times 10^{-5}$, $(1,2,\dots,9)\times 10^{-4}$, $(1,2,\dots,9)\times 10^{-3}$, and $(1.0,1.1,\dots,2.0)\times 10^{-2}$. For stars with initial masses $M \geq 11\,M_{\odot}$, we adopt models from our previous work \citet{Ou2023}. We first compute pre-main-sequence evolution to generate zero-age-main-sequence (ZAMS) models, which are then evolved to iron core collapse. For post-ZAMS evolution, we primarily use the default settings for massive stars from the \texttt{inlist\_massive\_defaults} in \texttt{MESAstar}, which include the following treatments. Convection is treated via mixing-length theory (MLT; \citealt{Henyey1965}) with $\alpha_{\rm MLT}= 1.5$, and the Ledoux criterion is applied to allow semi-convection with efficiency $\alpha_{\rm sc} = 0.01$. Thermohaline mixing follows \citet{Kippenhahn1980} with $\alpha_{\rm th} = 2$. We adopt the exponential overshooting scheme \citep{Herwig2000} with parameters: $f = 0.001$ in non-burning and H-burning regions, $f = 0$ in He- and metal-burning regions, and $f_0 = 0.0005$ in all regions. The "approx21 (Cr60)" nuclear network is employed, opacities use the "gs98" table with Type~2 enhancements for post-main-sequence C and O enrichment, and mass loss follows the prescriptions of \citet{Vink2001}, \citet{deJager1988}, and \citet{Nugis2000} for hot, cool, and Wolf-Rayet winds, respectively. Additional details can be found in \citet{Ou2023}.

For stars with $M \leq 10\,M_{\odot}$, we adopt the \texttt{MESA} test suite "1M\_pre\_ms\_to\_wd" to evolve models from the pre-main-sequence to the white dwarf stage. The nuclear reaction network "o18\_and\_ne22" includes 10 isotopes: $^1$H, $^3$He, $^4$He, $^{12}$C, $^{14}$N, $^{16}$O, $^{18}$O, $^{20}$Ne, $^{22}$Ne, and $^{24}$Mg. High- and low-temperature opacities are set using the "gs98" and "lowT\_fa05\_gs98" tables, respectively. Convection is treated with a mixing-length parameter $\alpha_{\rm MLT} = 2$. Mass loss follows \citet{Reimers1975} during the red giant branch and \citet{Bloecker1995} during the asymptotic giant branch phase.

From the grid, we select the $25\,M_{\odot}$, $Z=0.001$ model to perform numerical experiments on supergiant evolution. This model lies near the critical metallicity separating BSG and RSG outcomes \citep{Ou2023}, such that modest perturbations in stellar parameters can shift the evolutionary track between the two regimes. This sensitivity allows a direct probe of the envelope-expansion mechanism with minimal influence from other variables. In Section~\ref{sec:criterion},
we vary the triple-alpha reaction rate through a linear scaling factor, $\eta_{3\alpha}$, to examine how this modification influences envelope expansion and thereby identify the underlying expansion mechanism.

In Section~\ref{sec:envelope}, we conduct additional experiments on opacity by introducing the parameter $\zeta_{\kappa}$, which specifies the base metallicity ($Z_{\rm base}$ in MESA) used to retrieve opacity values. For the $25\,M_{\odot}$ models, we explore combinations of $Z$ and $\zeta_{\kappa}$, where $Z$ defines the star’s absolute metallicity and $\zeta_{\kappa}$ independently modulates the opacity.

\section{Physical Mechanism for Envelope Expansion}\label{sec:criterion}

\citet{Ou2023} showed that while some massive stars successfully evolve into RSGs during the core-He burning phase, others remain as BSGs without further expansion. Comparing these two evolutionary outcomes provides valuable insight into the criteria and physical mechanisms governing RSG formation. In this section, we analyze MESA models of $25\,M_{\odot}$ stars with a metallicity of $Z = 0.001$, varying only the triple-alpha reaction rate scaling factor, $\eta_{3\alpha}$, to identify the condition under which a star can evolve into an RSG.

\begin{figure}[tbh]
\centering
\includegraphics[scale=0.8]{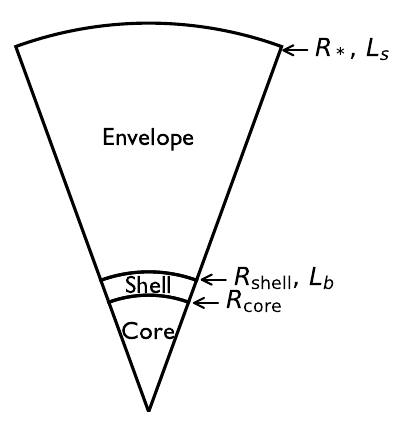}

\caption{Schematic diagram of the post-main-sequence stellar structure analyzed in this study, comprising a He core, an H-burning shell, and an envelope. The labeled radii and luminosities are: $R_{*}$ and $L_{s}$ at the stellar surface, $R_{\rm shell}$ and $L_{b}$ at the boundary between the envelope and the H-burning shell, and $R_{\rm core}$ at the interface between the shell and the He core.}
\label{fig:structure}
\end{figure}

For our analysis, we define three distinct stellar layers at the onset of the post-main-sequence stage, as illustrated schematically in Fig.~\ref{fig:structure}: the He core, the H-burning shell, and the envelope. The He core extends to a radius $R_{\rm core}$, defined as the location where the hydrogen mass fraction, $X_{\rm H}$, drops below 0.1, following MESA’s default convention. The boundary between the H-burning shell and the envelope, at radius $R_{\rm shell}$, is defined where the local energy generation rate from shell burning falls below $10^3~\rm erg~s^{-1}~g^{-1}$.

%%%%%%%%%%%%%%%%
%%%%%%%%%%%%%%%%

\subsection{Energy Budget}

One plausible explanation for RSG formation is that envelope expansion is powered by the gravitational energy released during core contraction. To examine this hypothesis and clarify the physical mechanism responsible for driving the transition to the RSG phase, we analyze the energy budgets associated with envelope expansion throughout the BSG and RSG stages.

We begin by defining the key quantities relevant to evaluating a star’s energy budget. In MESA outputs, the local energy generation rate is expressed as the sum of three components: the nuclear energy generation rate, $\epsilon_{\rm nuc}$; the neutrino loss rate, $\epsilon_{\nu}$; and the gravothermal term, $\epsilon_{\rm grav}$, which characterizes the local absorption or release of gravothermal energy. The gravothermal term is defined as
\begin{equation}
\epsilon_{\rm{grav}} \equiv -T \frac{\partial s}{\partial t},
\end{equation}
where $T$ is the temperature, $s$ is the specific entropy, and $t$ is the time. This term serves as a diagnostic for stellar expansion or contraction: $\epsilon_{\rm grav} < 0$ indicates expansion (heat absorption), while $\epsilon_{\rm grav} > 0$ indicates contraction (heat release).

The energy conservation equation in stellar interiors is written as \begin{equation} \frac{dl}{dm} = \epsilon_{\rm nuc} + \epsilon_{\rm grav} - \epsilon_{\nu}, \end{equation} 
where $l$ is the local luminosity and $m$ is the mass coordinate. Specifically, we define the integral of the gravothermal term as
\begin{equation}
\dot{Q} \equiv -\int_{M_1}^{M_2}\epsilon_{\rm{grav}} dm,
\end{equation}
with $M_1$ and $M_2$ denoting the lower and upper mass boundaries of the integration region. A positive $\dot{Q}$ implies net heat absorption, while a negative value indicates heat release. We denote the $\dot{Q}$ term over the envelope as $\dot{Q}_{\rm env}$ and over the core as $\dot{Q}_{\rm core}$.

By integrating Eq.~(2) along the stellar mass coordinates, we obtain:
\begin{equation} 
\Delta L = \dot{E}_{\rm nuc} - \dot{Q} - L_{\nu}, 
\end{equation} 
where $\dot{E}_{\rm nuc}$, $\dot{Q}$, and $L_{\nu}$ are the integrals of $\epsilon_{\rm nuc}$, $\epsilon_{\rm grav}$, and $\epsilon_{\nu}$, respectively. For the evolutionary stages of interest, $L_{\nu}$ are typically negligible, so Eq.~(4) simplifies to $\dot{Q} \simeq \dot{E}_{\rm{nuc}}-\Delta L$. Within the stellar envelope, there is virtually no nuclear reaction, so $\dot{E}_{\rm{nuc}}\simeq 0$. Therefore, the envelope's gravothermal term can be approximated by the thermal imbalance between the base and the surface:
\begin{equation}
\dot{Q}_{\rm env} \simeq -\Delta L = L_b - L_s,
\end{equation}
where $L_b$ and $L_s$ denote the luminosities at the base and surface of the envelope, respectively. This imbalance serves as a robust diagnostic of envelope expansion or contraction \citep[e.g.,][]{Renzini1992, Renzini1994, Ritossa1996}. During expansion, $L_b > L_s$ as radiative energy is partially trapped and used to perform $P$–$V$ work. In contrast, during contraction, $L_b < L_s$ as the gravothermal energy is released and radiated away.

\begin{figure*}[tbh]
\centering
\includegraphics[scale=0.45]{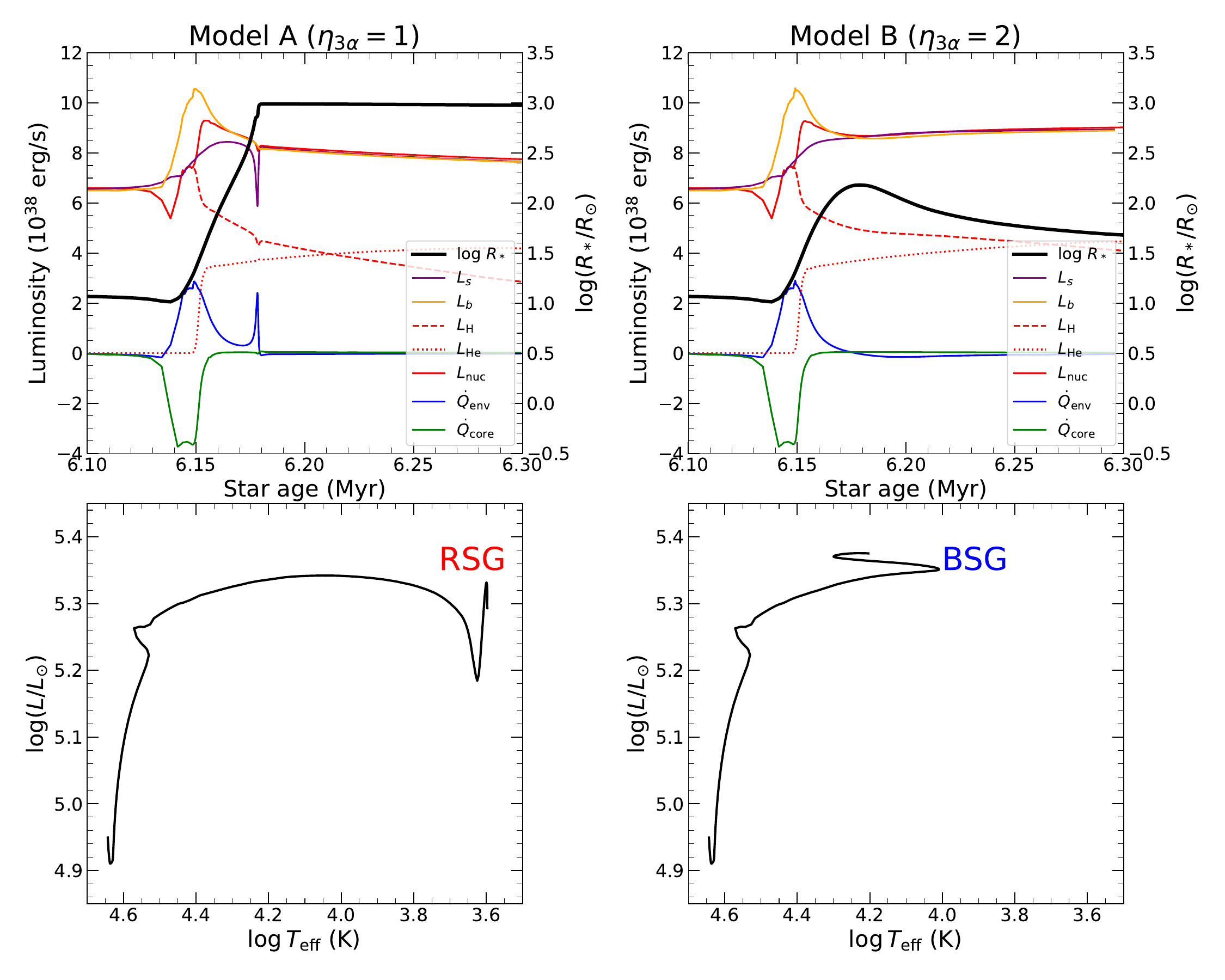}
\caption{Evolution of $25\,M_{\odot}$ stars at metallicity $Z=0.001$ with different He-burning rates, parameterized by a linear scaling factor $\eta_{3\alpha}$. Model A (\textit{left panels}), which adopts the standard rate ($\eta_{3\alpha}=1$), evolves into an RSG, whereas Model B (\textit{right panels}), with an enhanced rate ($\eta_{3\alpha}=2$), remains a BSG. The top panels show the temporal evolution of stellar radius ($R_*$), surface luminosity ($L_s$), luminosity at the base of the envelope ($L_b$), nuclear luminosities from H burning ($L_{\rm H}$) and He burning ($L_{\rm He}$), the total nuclear luminosity ($L_{\rm nuc}$), and the gravothermal heating rates of the core ($\dot{Q}_{\rm core}$) and envelope ($\dot{Q}_{\rm env}$). The bottom panels present the corresponding evolutionary tracks in the Hertzsprung-Russell diagram.}
\label{fig:energy}
\end{figure*}

In Fig.~\ref{fig:energy}, we compare the energy budgets of two MESA models with the same initial parameters—a mass of $25\,M_{\odot}$ and metallicity $Z=0.001$—but differing in the adopted value of $\eta_{3\alpha}$. Model A employs $\eta_{3\alpha} = 1$, corresponding to the standard triple-alpha reaction rate, and evolves into an RSG during the post-main-sequence phase. By contrast, Model B adopts $\eta_{3\alpha} = 2$ and remains a BSG throughout core He burning, without undergoing an RSG transition. For each model, we plot the evolution of various physical quantities in the top panels of Fig.~\ref{fig:energy} to analyze the energy budget: stellar radius ($R_*$), $L_s$, $L_b$, energy generation rate by H burning ($L_{\rm H}$), energy generation rate by He burning ($L_{\rm He}$), $\dot{Q}_{\rm core}$ and $\dot{Q}_{\rm env}$.

At the onset of the post-main-sequence phase, before core-He burning begins, Models A and B evolve in a similar manner. Around $\sim 6.13$ Myr, core contraction releases gravothermal energy, reflected by negative values of $\dot{Q}_{\rm core}$. Shortly thereafter, shell-H burning is ignited, and the envelope begins to expand, leading to an increase in the stellar radius $R_*$. The expansion is powered by both the gravothermal energy released from the contracting core and the energy generated by the H-burning shell.

The evolutionary paths of the two models diverge after core-He burning commences. Model A continues to expand into an RSG, reaching a radius of $\sim 1,000\,R_{\odot}$, while Model B halts its expansion at $\sim 150\,R_{\odot}$ and remains in the BSG phase. As shown in the HR diagrams in the lower panels of Fig.~\ref{fig:energy}, the surface temperature of Model A steadily decreases to $\sim 4,000$ K, characteristic of an RSG, whereas Model B cools only to $\sim 10,000$ K before reheating, never entering the RSG regime.

If the envelope expansion in RSGs were governed solely by the energy released from core contraction, one would expect Model A’s core to release more energy than that of Model B, since its envelope expands much further. However, Fig.~\ref{fig:energy} shows that this is not the case: the evolution of $\dot{Q}_{\rm core}$ is nearly identical in both models, exhibiting a negative peak at $\sim$6.13–6.16 Myr before returning to values near zero. After $\sim$6.16 Myr, although the envelope of Model A continues to expand, its core ceases to release significant gravothermal energy. At this stage, the positive $\dot{Q}_{\rm env}$ that drives the expansion in Model A is primarily contributed by the decrease in $L_s$ rather than an increase in $L_b$. In fact, $L_b$ in Model A decreases even more rapidly than in Model B. Notably, during 6.17--6.18 Myr, Model A exhibits a prominent bump in $\dot{Q}_{\rm env}$, which is not accompanied by any increase in $|\dot{Q}_{\rm core}|$, but corresponding to the sharp drop in $L_s$—the key difference from Model B. This indicates that the further envelope expansion toward the RSG phase in Model A is not driven by additional energy input from the core, but rather by a larger fraction of the outgoing luminosity being trapped within the envelope. Thus, core gravothermal energy release alone cannot account for RSG formation.

Therefore, this example demonstrates that the evolution toward an RSG can be divided into two distinct stages, characterized by the energy budget of the stellar envelope:

\paragraph{(1) From TAMS to BSG.} Both models undergo this stage, expanding to $\sim 100\,R_{\odot}$ as BSGs. During this phase, core contraction releases gravothermal energy, reflected by negative $\dot{Q}_{\rm core}$. The envelope responds by absorbing energy from both core contraction and shell-H burning, producing a positive peak in $\dot{Q}_{\rm env}$.

\paragraph{(2) From BSG to RSG.} At the BSG phase, the evolutionary paths of the two models diverge. While Model B halts its expansion and remains a BSG, Model A continues to expand into the RSG regime as its envelope increasingly traps outgoing radiation. This second stage of expansion is not powered by gravothermal energy from the core and thus is not an inevitable consequence of core contraction.

%%%%%%%%%%%%%%%%
%%%%%%%%%%%%%%%%

\subsection{The Mirror Principle}

\begin{figure*}[tbh]
\centering
\includegraphics[scale=0.45]{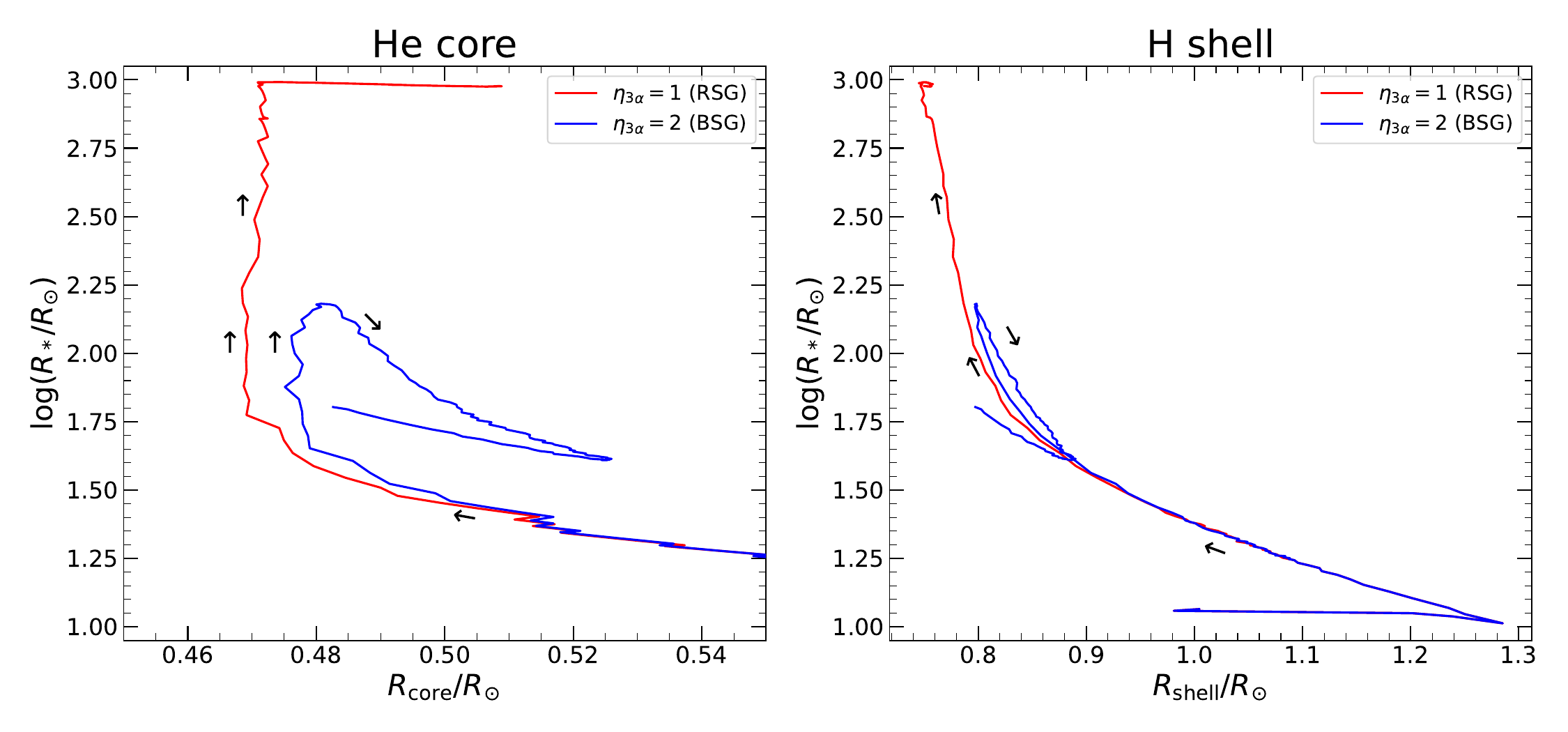}
\caption{Evolution of the stellar radius ($R_*$) alongside the He core radius ($R_{\rm core}$) and the radius at the outer edge of the H-burning shell ($R_{\rm shell}$). Arrows indicate the direction of time progression during stellar evolution. Notably, $R_*$ generally evolves inversely to $R_{\rm shell}$.}
\label{fig:Rcore}
\end{figure*}

While the scenario in which core energy release drives RSG formation is ruled out, we next examine whether the conventional “mirror principle” holds. In the left panel of Fig.~\ref{fig:Rcore}, 
we show the evolution of $R_*$ alongside $R_{\rm core}$ for Models A and B to assess how the envelope responds to core contraction. At the onset of the post-main-sequence phase, both models evolve similarly: as the He core contracts, the envelope expands. The envelope thus appears to vary inversely with the core—core contraction leads to envelope expansion, whereas core expansion halts or reverses it.

Once the core contraction halts at $R_* \sim 100\,R_{\odot}$, however, the envelope evolution of two models diverge. In Model B, the envelope halts its expansion at $\sim 150\,R_{\odot}$ and gradually contracts to $\sim 40\,R_{\odot}$. In Model A, by contrast, the envelope continues expanding from $\sim 100\,R_{\odot}$ to $\sim 1,000\,R_{\odot}$, while the core remains nearly constant radius of at $R_{\rm core} \sim 0.47\,R_{\odot}$. This indicates that continued envelope expansion of Model A from the BSG phase toward the RSG phase is not driven by core contraction. This conclusion is consistent with the energy budget in Fig.~\ref{fig:energy}, where gravothermal energy input from the core ($\dot{Q}_{\rm core}$) is negligible during the BSG-to-RSG transition. Therefore, the traditional “mirror principle” based on core contraction fails to explain RSG formation in this model.

%%%%%%%%%%%%%%%%
%%%%%%%%%%%%%%%%

A more robust correlation emerges when $R_{\rm core}$ is replaced by the radius at the outer edge of the H-burning shell, $R_{\rm shell}$, as shown in the right panel of Fig.~\ref{fig:Rcore}. In both models, the envelope’s outer boundary (traced by $R_*$) and the shell-envelope boundary (traced by $R_{\rm shell}$) consistently move in opposite directions. Initially, the burning shell contracts inward while the envelope expands. After the evolutionary paths of the two models diverge at $R_*\gtrsim 100\,R_{\odot}$, this inverse relationship persists: in Model A, $R_{\rm shell}$ continues moving inward as $R_*$ increases to $\sim 1,000\,R_{\odot}$; in Model B, $R_{\rm shell}$ reverses outward at $R_* \sim 150\,R_{\odot}$, coinciding with the onset of envelope contraction. Throughout the evolution, changes in $R_{\rm shell}$ and $R_*$ remain consistently anti-correlated.

This result refines the traditional “mirror principle” for stars with a burning shell. Rather than an inverse relationship between the envelope and the core, the more accurate description is that the inner and outer boundaries of the envelope evolve in opposite directions. This principle demonstrated using the $25\,{M_{\odot}}$ models is generalized to different masses and metallicities in Section~\ref{sec:mass}, while its underlying physical mechanism is investigated in Section~\ref{sec:mirror} and in a companion Paper II.

\subsection{Role of Envelope Base Luminosity}

Under the refined mirror principle, the difference in supergiant outcomes between Models A and B can be reduced to a single question: what causes the bifurcation the evolution of $R_{\rm shell}$ between them? The key is to examine the turning point of $R_{\rm shell}$ in Model B and identify what triggers this reversal, which prevents further expansion toward the RSG regime.

\begin{figure}[tbh]
\centering
\includegraphics[scale=0.4]{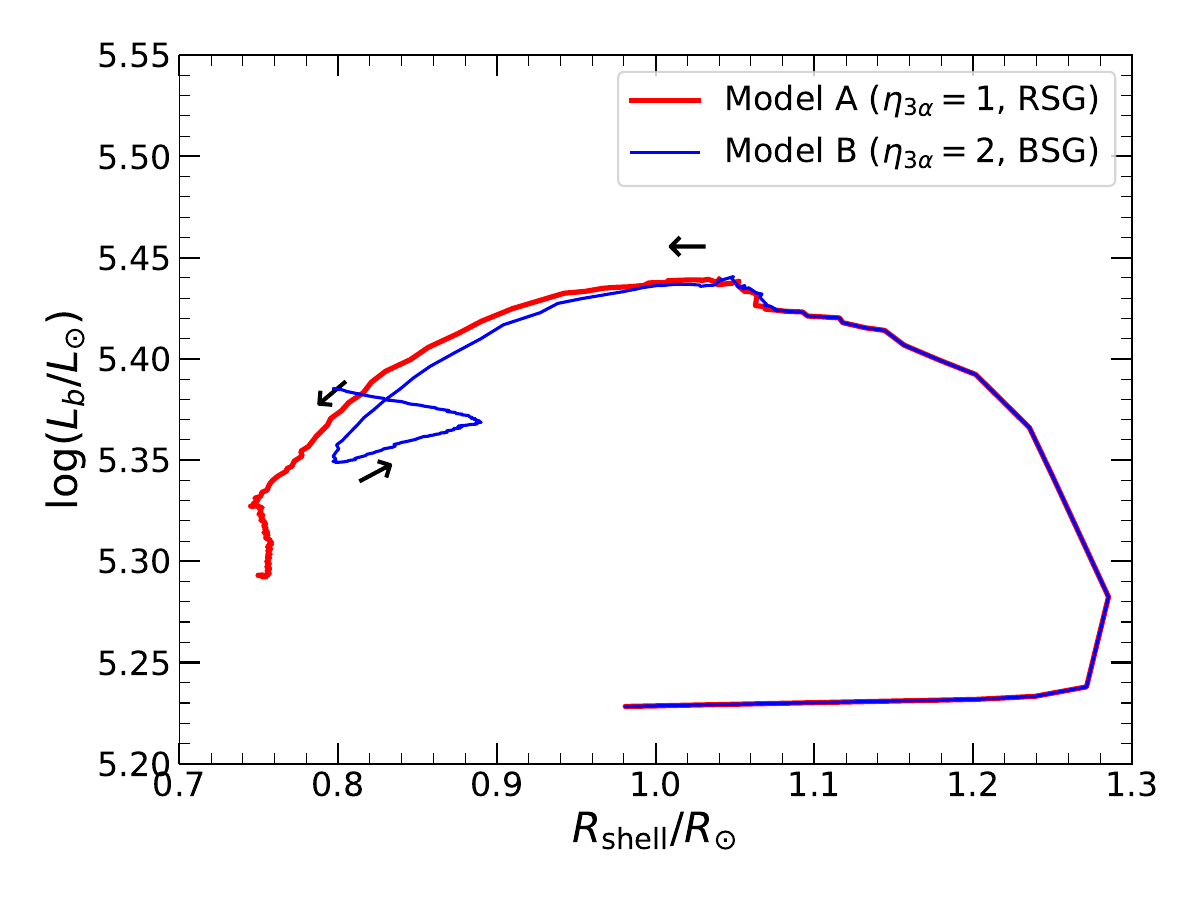}
\caption{Evolution of the radius and luminosity at the interface between the H-burning shell and the envelope for Models A and B. The radius and luminosity at the shell-envelope boundary are denoted as $R_{\rm shell}$ and $L_b$, respectively. Arrows indicate the direction of temporal evolution.}
\label{fig:Rshell_Lb}
\end{figure}

We find that the key factor associated with this turning point is the local luminosity at the radius of $R_{\rm shell}$, denoted as $L_b$. As illustrated in Fig.~\ref{fig:Rshell_Lb}, the reversal of $R_{\rm shell}$ in Model B from inward to outward motion coincides precisely with the point at which $L_b$ transitions from decreasing to increasing. Combining with the relationship shown in Fig.~\ref{fig:Rcore}, the turning points of $L_b$, $R_{\rm shell}$, and $R_*$ all occur simultaneously. In contrast, in Model A, both $L_b$ and $R_{\rm shell}$ continue decreasing until $R_{\rm shell}$ reaches $\sim 0.75\,R_{\odot}$.

At this turning point, core contraction has already ceased (Fig.~\ref{fig:Rcore}), so the luminosity at the base of the envelope is primarily supplied by nuclear burning, including shell-H and core-He burning. Consequently, $L_b$ closely approximates the total nuclear energy generation rate, $L_{\rm nuc}$. Given the established correlation between $L_b$ and $R_{\rm shell}$, and the anti-correlation between $R_{\rm shell}$ and $R_*$, an anti-correlation between $L_{\rm nuc}$ and $R_*$ is expected.

\begin{figure}[tbh]
\centering
\includegraphics[scale=0.5]{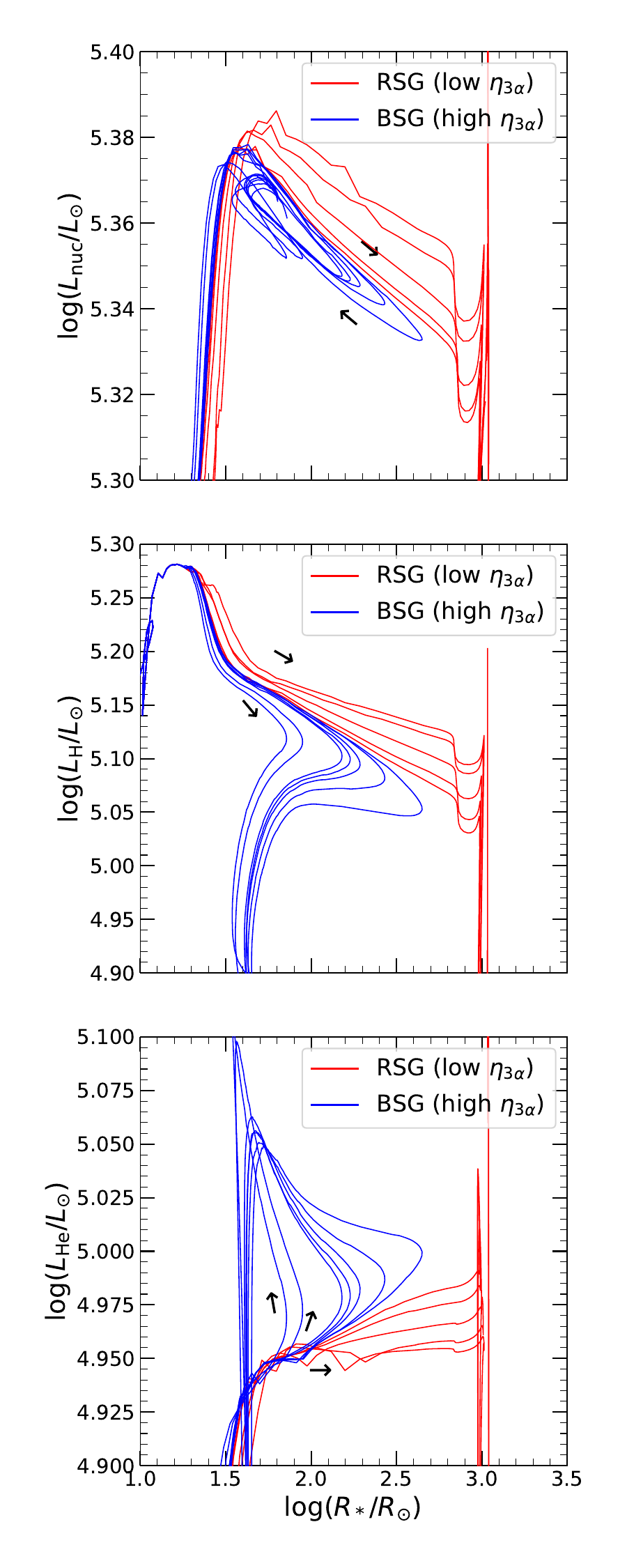}
\caption{Evolution of the energy generation rates as a function of stellar radius ($R_*$) for $25\,M_{\odot}$, $Z = 0.001$ models with different values of $\eta_{3\alpha}$. The three panels show the total nuclear energy generation rate ($L_{\rm nuc}$), the contribution from hydrogen burning ($L_{\rm H}$), and from helium burning ($L_{\rm He}$). Red curves represent models with lower $\eta_{3\alpha}$ (0.1, 0.2, 0.5, 1.0, and 1.2) that eventually evolve into RSGs, while blue curves represent models with higher $\eta_{3\alpha}$ (1.33, 1.4, 1.6, 1.8, 2.0, 5.0, 10) that contract after minor expansion and remain as BSGs.}
\label{fig:luminosity}
\end{figure}

Fig.~\ref{fig:luminosity} illustrates how $L_{\rm nuc}$, $L_{\rm H}$, and $L_{\rm He}$ vary with $R_*$ in $25\,M_{\odot}$, $Z = 0.001$ models computed with different values of $\eta_{3\alpha}$. For models with smaller $\eta_{3\alpha}$ (red curves), $L_{\rm nuc}$ steadily decreases after reaching its maximum, while $R_*$ continues to grow, ultimately reaching the RSG regime at $\sim 1,000\,R_{\odot}$. In contrast, models with larger $\eta_{3\alpha}$ (blue curves) show a subsequent rise in $L_{\rm nuc}$. Remarkably, the turning point of $L_{\rm nuc}$ in each model coincides precisely with that of $R_*$, revealing a strong correlation between the nuclear energy generation rate and the envelope’s evolution. This correlation arises through the coupling between $R_{\rm shell}$ and $R_*$, referred to as the refined mirror principle.

Examining the individual contributions of shell H burning ($L_{\rm H}$) and core He burning ($L_{\rm He}$) to $L_{\rm nuc}$ reveals the mechanism behind this behavior. In models with larger $\eta_{3\alpha}$ (blue curves), although $L_{\rm H}$ declines earlier at smaller $R_*$, $L_{\rm He}$ rises much more sharply at the same time, compensating for the decrease in $L_{\rm H}$ and leading to an overall increase in $L_{\rm nuc}$. This rise in $L_{\rm nuc}$ drives the outward motion of $R_{\rm shell}$ and triggers envelope contraction. Conversely, in models with smaller $\eta_{3\alpha}$, $L_{\rm He}$ increases more gradually and does not offset the decline in $L_{\rm H}$, resulting in a continued decrease in $L_{\rm nuc}$, sustained envelope expansion, and eventual evolution into the RSG phase. For these models, the criterion for the continued envelope expansion after the core contraction halts is expressed as:
\begin{equation}
    \frac{dL_{\rm nuc}}{dt} = \frac{d}{dt}(L_{\rm H}+L_{\rm He}) < 0.
\end{equation}
As long as this condition is satisfied, the envelope continues to expand. Once $dL_{\rm nuc}/dt > 0$, the stellar envelope of the BSG begins to stabilize immediately.

%%%%%%%%%%%%%%%%
%%%%%%%%%%%%%%%%
%%%%%%%%%%%%%%%%
%%%%%%%%%%%%%%%%

As the experiments above involve adjusting the nuclear burning rates through $\eta_{3\alpha}$, it is more reasonable to regard the central nuclear burning as the cause and the envelope expansion as the response, rather than the reverse. In short, once the He core ceases contracting, nuclear burning in the central region regulates the location of the H-burning shell and, consequently, the envelope radius.

Importantly, the mechanism by which the central regions of a star govern envelope expansion toward the RSG phase is not simply a matter of energy supply. From the TAMS stage to the BSG stage, while the core is still contracting, an enhanced energy flux at the base of the envelope indeed corresponds to envelope expansion. However, once core contraction ceases, further expansion requires a decline—not an increase—in $L_b$ (or equivalently, $L_{\rm nuc}$). If $L_{\rm He}$ rises too rapidly, it can drive $L_{\rm nuc}$ upward and instead triggering envelope contraction. Thus, the energy-supply scenario fails to account for RSG formation, contradicting earlier claims that increasing core luminosity directly promotes expansion \citep[e.g.,][]{Renzini1994}.

The envelope’s expansion or contraction depends not directly on the amount of energy it receives but on the required structural configuration, governed by the motion of the burning shell according to the refined mirror principle. Even as the energy flux from the interior decreases, as long as $R_{\rm shell}$ is moving inward, the supergiant envelope continues to expand by trapping a larger fraction of outgoing radiation, reflecting a decrease in surface luminosity $L_s$.

\section{General Criteria for Stars to Turn Red}\label{sec:mass}

In the previous section, we used the experiments of $25\,M_{\odot}$ stars to rule out the scenario of energy supply for explaining the RSG formation and established the refined mirror principle.
In this section, we examine stellar models spanning a broad range of masses and metallicities to obtain general criteria for stars to turn red.

\subsection{Evolution Across Different Masses and Metallicities}

To generalize the criteria for RSG formation, we analyze our grid of stellar models with initial masses spanning $1$–$30\,M_{\odot}$ and metallicities ranging from $Z = 10^{-5}$ to $2 \times 10^{-2}$, as described in Section~2.1.

\subsubsection{Examples of $5$, $15$, and $25\,M_{\odot}$}
\begin{figure*}[tbh]
\centering
\includegraphics[width=\textwidth]{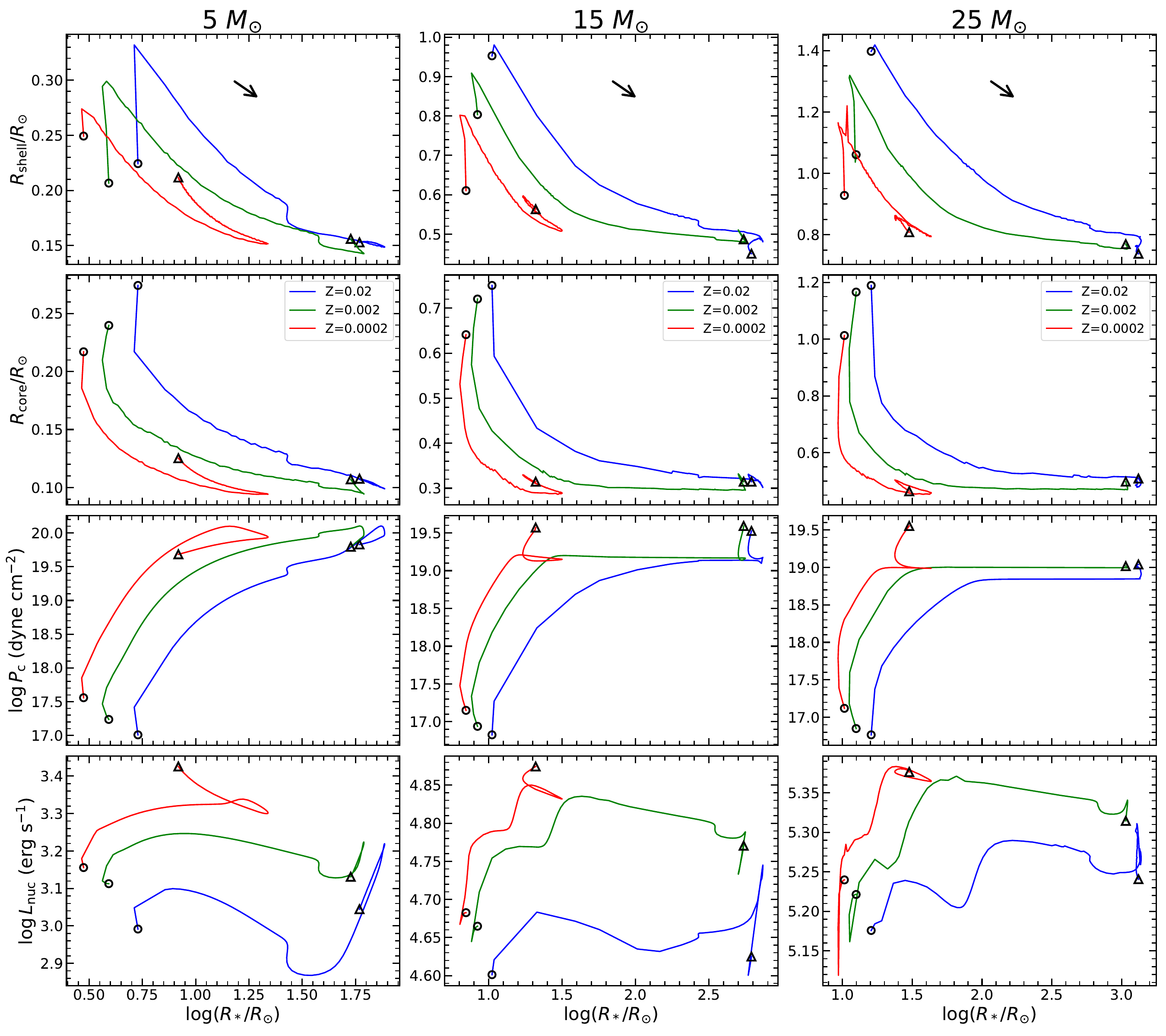}
\caption{Evolution of the shell radius ($R_{\rm shell}$), core radius ($R_{\rm core}$), central pressure ($P_c$), and nuclear luminosity ($\log L_{\rm nuc}$) as functions of the stellar radius ($R_*$) during the post-main-sequence stage, prior to carbon core formation, for $5$, $15$, and $25\,M_{\odot}$ stars at different metallicities. Circles and triangles mark the formation of the He core and C core, respectively, defining the start and end of the time span shown. Arrows in the top row indicate the general direction of evolution in time, with $R_{\rm shell}$ decreasing and $R_*$ increasing.}
\label{fig:RSG_examples_M_Z}
\end{figure*}

As illustrative examples, we first examine the evolution of three stellar masses: $5\,M_{\odot}$, $15\,M_{\odot}$, and $25\,M_{\odot}$. For each mass, we consider three representative metallicities: $Z = 0.02$ (high), $0.002$ (intermediate), and $0.0002$ (low). Fig.~\ref{fig:RSG_examples_M_Z} shows the post-main-sequence evolution of $R_{\rm shell}$, $R_{\rm core}$, central pressure ($P_{\rm c}$), and $L_{\rm nuc}$ prior to carbon core formation for each case. 

First, we verify the refined mirror principle established in Section~\ref{sec:criterion}. 
In all examples, the stellar radius $R_*$ generally evolves in the opposite direction of $R_{\rm shell}$ throughout the post-main-sequence phase. For the low-metallicity ($Z=0.0002$) models that fail to evolve into the RG/RSG phase, the turning points of $R_*$ coincide with those of $R_{\rm shell}$. These evolutionary tracks are therefore consistent with the refined mirror principle.

Next, we evaluate the applicability of the criterion for continued envelope expansion at the BSG stage (Eq. 6). In the low-metallicity ($Z = 0.0002$) models that cease expanding at the BSG stage, the turning points of $R_*$ all match those of $L_{\rm nuc}$, indicating that Eq. (6) holds. For the intermediate-metallicity ($Z=0.002$) and high-metallicity ($Z=0.02$) models that successfully turn red, the situations are more complicated as the evolution depends on whether the core contraction is still ongoing.

\begin{figure*}[tbh]
\centering
\includegraphics[scale=0.5]{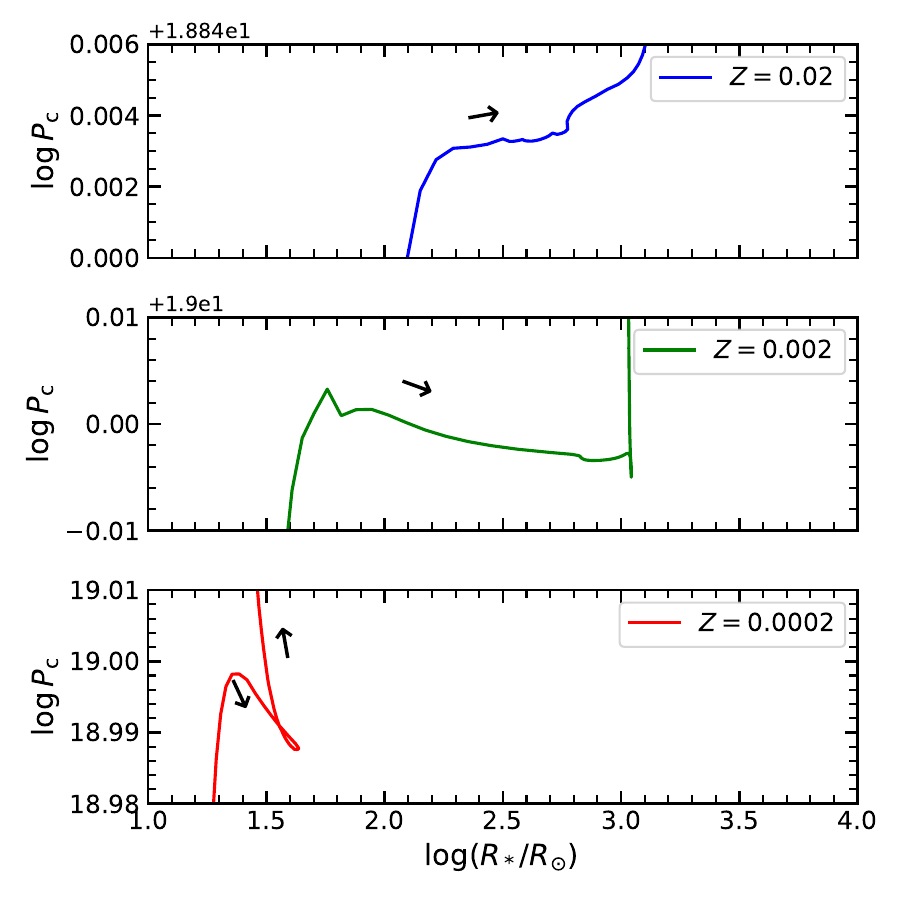}
\caption{Evolution of the central pressure ($P_{\rm c}$, in cgs units) for $25\,M_{\odot}$ stars at different metallicities. Arrows indicate the direction of temporal evolution.}
\label{fig:cntrP}
\end{figure*}

To determine when core contraction ends, using $R_{\rm core}$ can be ambiguous in some cases because it traces the location defined by $X_{\rm H} = 0.1$, rather than reflecting the actual central condition of the star. Instead, we adopt the central pressure, $P_{\rm c}$, as a more robust diagnostic of core contraction. Fig.~\ref{fig:cntrP} presents a zoom-in of $P_{\rm c}$ versus $R_*$ for the $25\,M_{\odot}$ models. For $Z=0.002$, $P_{\rm c}$ reaches a maximum at $\log(R_*/R_{\odot}) \sim 1.75$ before declining, whereas for $Z=0.02$, $P_{\rm c}$ continues to increase even as the star expands into an RSG with $\log(R_*/R_{\odot}) > 3$. These results indicate that although both models evolve into the RSG phase, the former turns red while core contraction is still ongoing, whereas in the latter, core contraction is complete before the star becomes red.

According to these models, there are two channels through which stars can turn red. While Eq. (6) applies once core contraction has ceased, there exists another channel in which stars evolve directly into the RG/RSG phase during core contraction. The two channels are:

\paragraph{(1) Stars that turns red during core contraction.} In cases where $P_{\rm c}$ continues to rise during envelope expansion toward the RG/RSG phase, $R_{\rm shell}$ moves inward together with the core, and the envelope’s growth into the RSG regime proceeds independently of the evolution of $L_{\rm nuc}$. Four of the nine models in Fig.~\ref{fig:RSG_examples_M_Z} fall into this category—all three high-metallicity models ($Z = 0.02$) and the $5\,M_{\odot}$, $Z = 0.002$ model.

\paragraph{(2) Stars that turn red after core contraction ceases.} For this case, Eq. (6) provides a reliable criterion for determining continued envelope expansion continues. In the intermediate-metallicity ($Z = 0.002$) models of $15\,M_{\odot}$ and $25\,M_{\odot}$, $L_{\rm nuc}$ steadily declines after core contraction ends, allowing the envelope to expand continuously into the RG/RSG regime.

We note that all the evolutionary criteria described above are grounded in the refined mirror principle.

%%%%%%%%%%%%%%%%
%%%%%%%%%%%%%%%%
%%%%%%%%%%%%%%%%
%%%%%%%%%%%%%%%%

\subsubsection{Relationship Across the Full Grid of Models}

To verify the universality of the expansion criteria, we examine our full grid of stellar models spanning a broad range of initial masses and metallicities. Among the 1,140 models with initial masses of $1$–$30\,M_{\odot}$ and various metallicities, we exclude 38 low-metallicity cases (mostly with $24$–$30\,M_{\odot}$) in which the C core forms before the He core has completed contraction. These cases fall outside the scope of this study, yielding a total of 1,102 models for analysis.

\begin{figure*}[tbh]
\centering
\includegraphics[width=\textwidth]{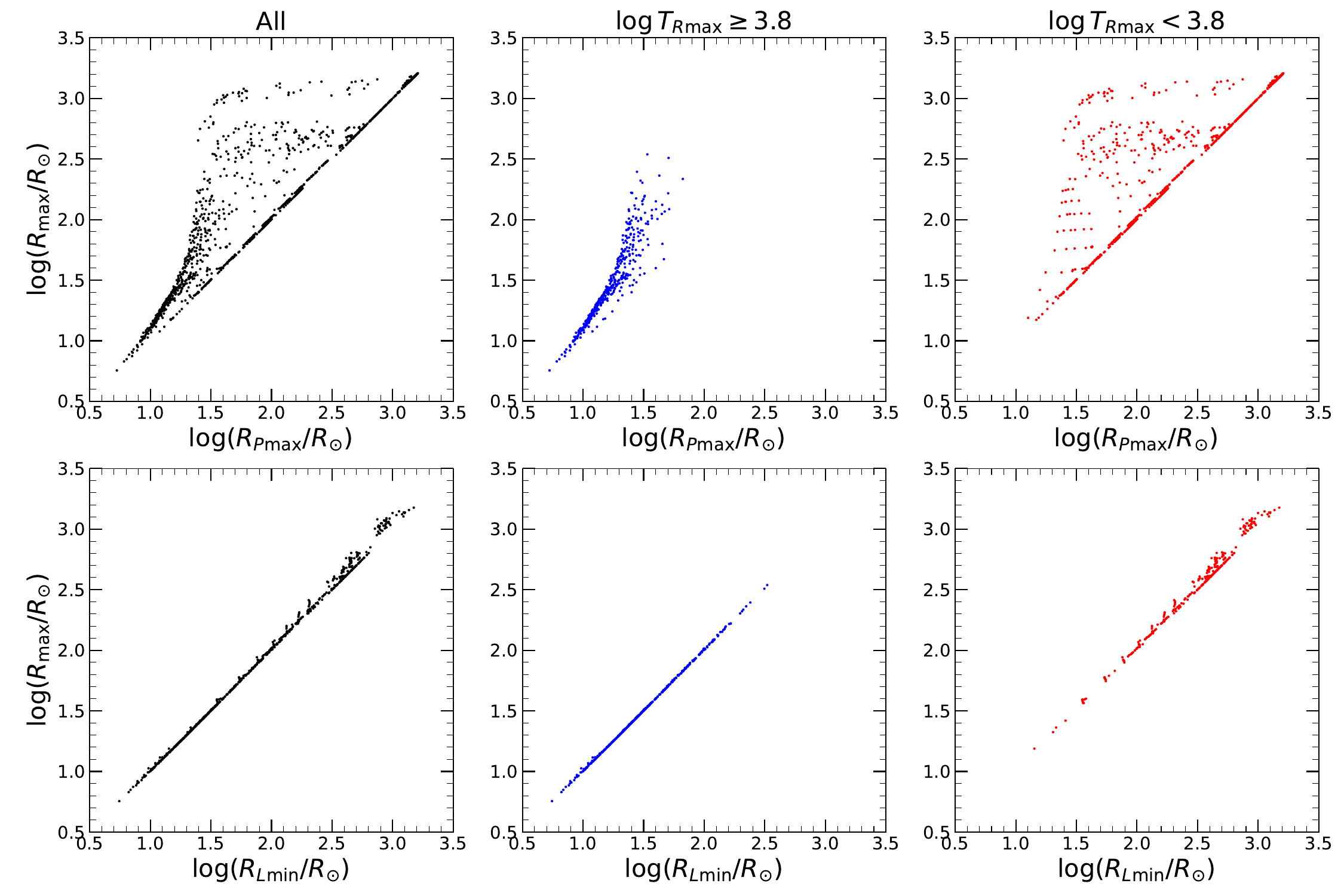}
\caption{Scatter plots of characteristic radii during core-He burning for stellar models with different masses and metallicities ($1$–$30\,M_{\odot}$, $Z = 10^{-5}$–$2\times10^{-2}$). Top: maximum stellar radius ($R_{\rm max}$) versus the radius of peak central pressure ($R_{P\rm max}$) at the end of core contraction for all the 1,102 models. Bottom: $R_{\rm max}$ versus the radius where nuclear energy generation reaches its minimum after core contraction ($R_{L\rm min}$) for the 711 models with $R_{\rm max} > R_{P{\rm max}}$. Left: all models; middle: BG/BSG ($\log T_{R\rm max} \ge 3.8$); right: RG/RSG ($\log T_{R\rm max} < 3.8$).}
\label{fig:RLmin}
\end{figure*}

For each model, we focus on the evolutionary period after He-core formation but before C-core development. Within this interval, we identify the maximum stellar radius, $R_{\rm max}$. We then examine the central pressure, $P_{\rm c}$, and locate its local maximum during core-He burning, with the corresponding stellar radius denoted as $R_{P{\rm max}}$. The top panels of Fig.~\ref{fig:RLmin} present scatter plots of $R_{P{\rm max}}$ versus $R_{\rm max}$. Models are classified as red or blue according to the effective temperature at $R_{\rm max}$, $T_{R{\rm max}}$, adopting $\log (T_{R{\rm max}}/{\rm{K}}) < 3.8$ (i.e., $T_{R{\rm max}} \lesssim 6{,}300$\,K) as the threshold for RG/RSG identification.

In 391 RG/RSG models, $R_{P{\rm max}}$ coincides with $R_{\rm max}$, indicating that the envelope attains its maximum radius while $P_{\rm c}$ is still increasing—that is, while the core is still contracting. These stars expand directly to the RG/RSG extent during core contraction.

For the remaining 711 models, $R_{\rm max} > R_{P{\rm max}}$, indicating that the envelope continues to expand after core contraction ceases. For these stars, we identify the minimum nuclear energy-generation rate, $L_{\rm nuc}$, occurring after the peak of $P_{\rm c}$, and denote the corresponding stellar radius as $R_{L{\rm min}}$. The lower panels of Fig.~\ref{fig:RLmin} display $R_{L{\rm min}}$ versus $R_{\rm max}$ for these 711 models with $R_{\rm max} > R_{P{\rm max}}$. In particular, in all BG/BSG cases, $R_{L{\rm min}}$ and $R_{\rm max}$ are nearly identical, indicating that the turning points of $L_{\rm nuc}$ coincide with those of $R_*$—consistent with the relationships shown in Figs.~\ref{fig:luminosity} and \ref{fig:RSG_examples_M_Z}. Similarly, in the RG/RSG cases, although some scatter is present, $R_{L{\rm min}}$ still closely aligns with $R_{\rm max}$, showing that $L_{\rm nuc}$ continues to decline throughout the expansion until the envelope reaches its final size of RG/RSG. These results confirm that Eq.~(6) provides a robust and general criterion for continued envelope expansion after core contraction, applicable across a broad range of stellar masses and metallicities.

%%%%%%%%%%%%%%%%
%%%%%%%%%%%%%%%%

\subsection{Roadmap Toward Red Giants or Supergiants}

\begin{figure*}[tbh]
\centering
\includegraphics[scale=0.5]{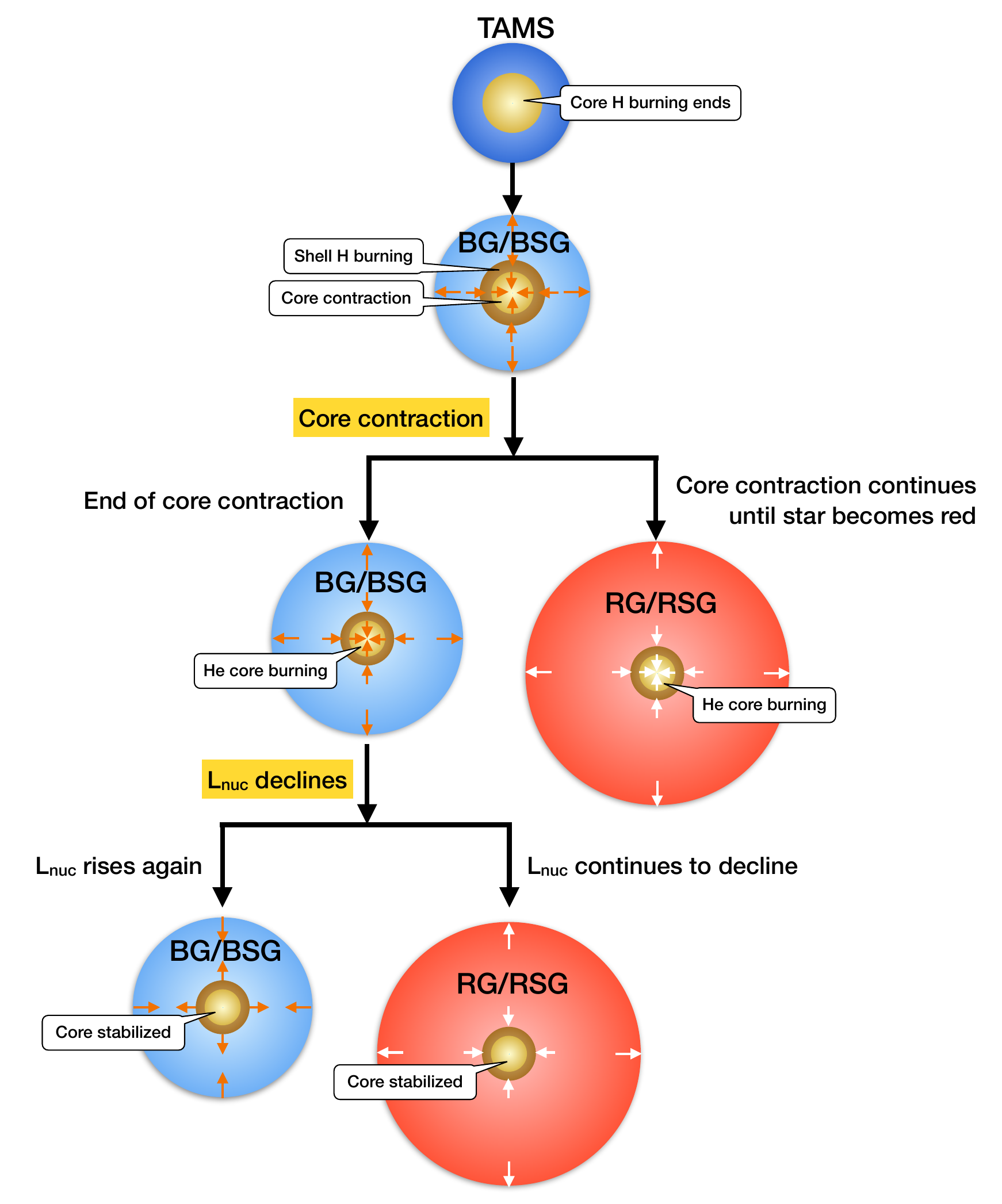}
\caption{Flow chart illustrating the evolutionary pathways toward the RG/RSG phase.}
\label{fig:flow}
\end{figure*}

We have verified the general mechanism and criteria for a star to evolve into an RG/RSG, as summarized in Fig.~\ref{fig:flow}. At the TAMS, core H burning ends and the He core contracts, forming an H-burning shell that moves inwards with the core. The envelope expands in the opposite direction of the inward-moving shell. The extent of this expansion by the end of core contraction depends on mass and metallicity. In some cases, the star reaches RG/RSG dimensions while the He core is still contracting and the central pressure continues to rise—representing the first evolutionary channel toward RG/RSG formation.

If the core completes contraction and stabilizes at the blue giant (BG) or BSG phase, further envelope expansion depends on the nuclear energy generation rate from core He burning and shell H burning. At the envelope base, the local luminosity ($L_b \simeq L_{\rm nuc}$) and radius ($R_{\rm shell}$) are tightly correlated. As $L_{\rm nuc}$ declines, $R_{\rm shell}$ moves inward and the envelope continues expanding. Conversely, if $L_{\rm nuc}$ rises during the BSG stage—e.g., due to rapid $L_{\rm He}$ growth—$R_{\rm shell}$ reverses outward, the envelope contracts, and the star remains a BG/BSG. If $L_{\rm nuc}$ continues to decrease, $R_{\rm shell}$ keeps moving inward, allowing expansion into the RG/RSG regime. After core contraction, the net energy generation rate thus dictates the star’s evolutionary path.

We have summarized the evolutionary pathways toward supergiants and the corresponding criteria. In this paper, we do not address what determines which pathway a star follows; this question is further explored in \citet{Ou2025} in the context of explaining the critical metallicity for RSG formation.

The underlying mechanism for all these pathways follows the refined mirror principle, in which the inner and outer boundaries of the envelope evolve in opposite directions. Although this principle is currently derived empirically from our models, the next section and Paper~II \citep{paper2} explore its physical origin.

%%%%%%%%%%%%%%%%
%%%%%%%%%%%%%%%%

\section{Physics of the Refined Mirror Principle}\label{sec:mirror}

We refine the “mirror principle” for stars with an H-burning shell: the envelope expands or contracts opposite to the motion of its inner boundary, defined by the outer edge of the H-burning shell rather than the He core. The physical foundation of this principle lies in a key feature of this shell-envelope boundary-- its nearly constant temperature and pressure.

\begin{figure*}[tbh]
\centering
\includegraphics[scale=0.4]{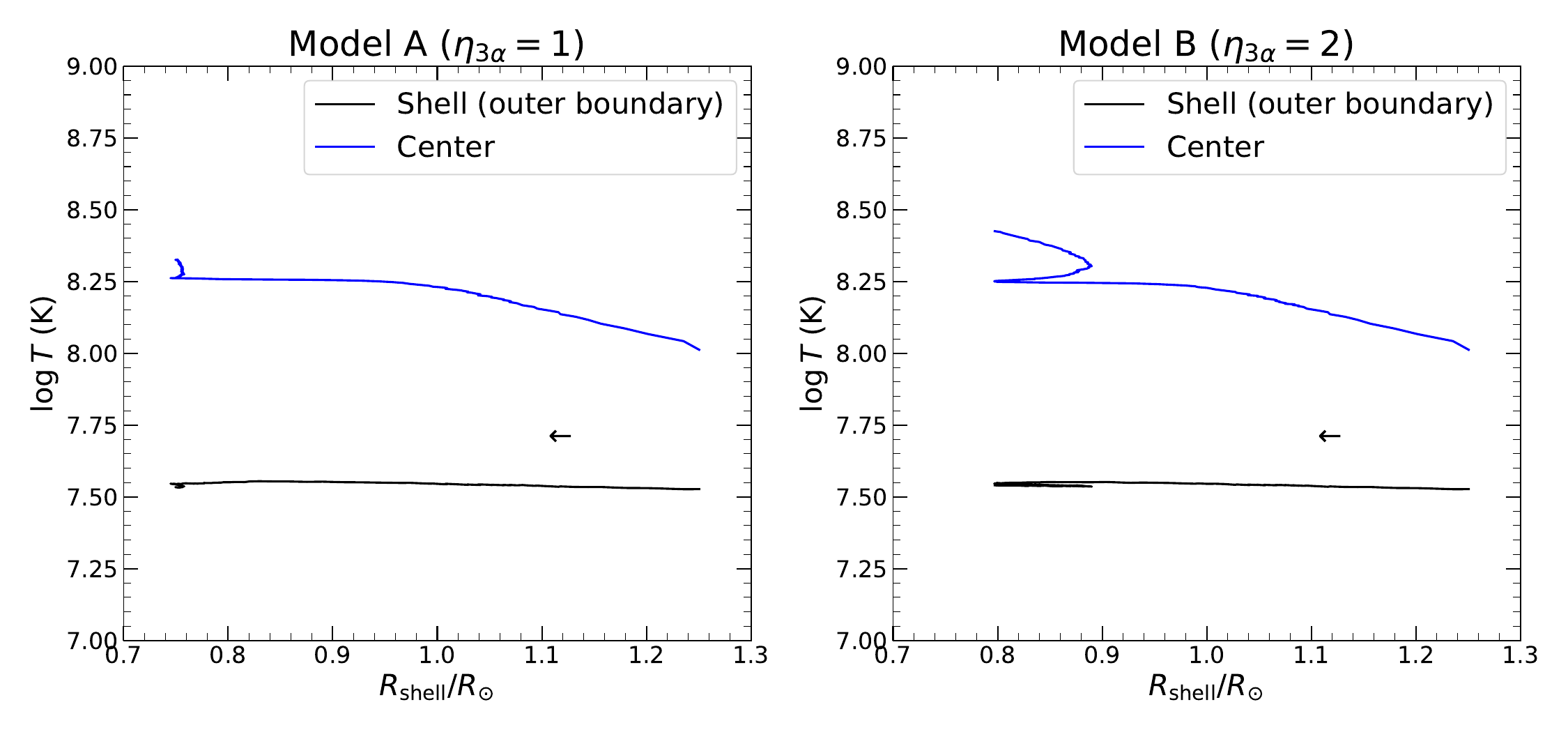}
\includegraphics[scale=0.4]{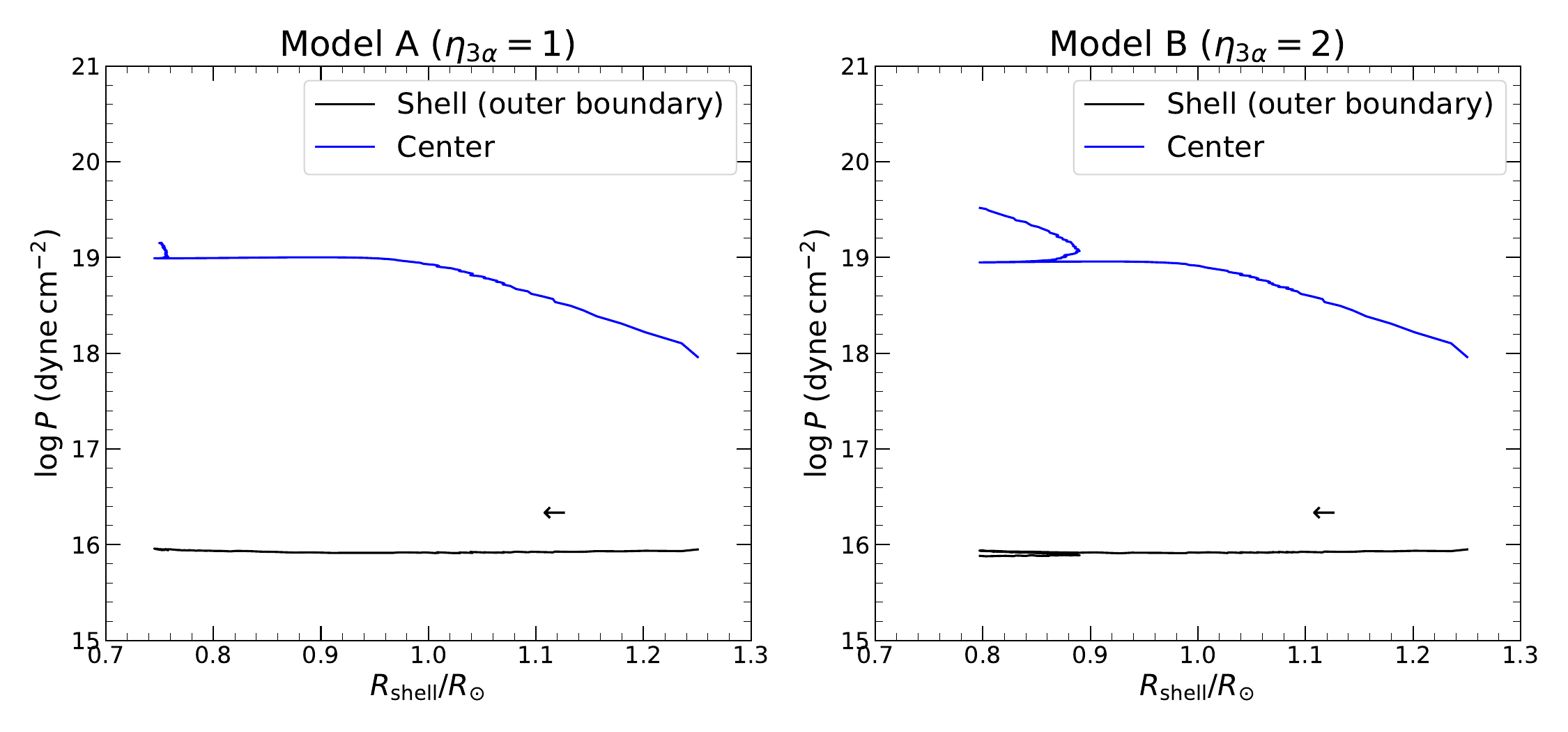}
\caption{Temperature and pressure evolution at the center and at the shell-envelope boundary for Models A and B. While central values change, the shell-envelope boundary maintains nearly constant temperature and pressure throughout, with arrows indicating the direction of evolution.}
\label{fig:Tshell}
\end{figure*}

As shown in the top panels of Fig.~\ref{fig:Tshell}, while $R_{\rm shell}$ and the central temperature continues continue to evolve, the temperature at the shell-envelope boundary (the location of $R_{\rm shell}$) remain nearly constant at $\sim (3-4)\times 10^7\,$K, i.e., $\log (T/\rm{K}) \sim 7.5-7.6$. Meanwhile, the bottom panels of Fig.~\ref{fig:Tshell} show that the pressure at the shell-envelope boundary remains almost constant.

The physical reason for the nearly constant temperature at the shell-envelope boundary is that this boundary is set by the ignition threshold of the CNO cycle, typically at temperatures of a few times $10^7$ K. Meanwhile, the $P$–$T$ stratification in the central region of the star generally follows the adiabatic temperature gradient, which remains nearly constant ($\sim$0.25 in these $25\,M_{\odot}$ models), causing the pressure at the shell-envelope boundary to remain nearly constant as well. Thus, the base of the stellar envelope is characterized by an almost constant temperature and pressure, rather than by a fixed mass or radius coordinate.

%%%%%%%%%%%%%%%%
%%%%%%%%%%%%%%%%

\begin{figure*}[tbh]
\centering
\includegraphics[scale=0.45]{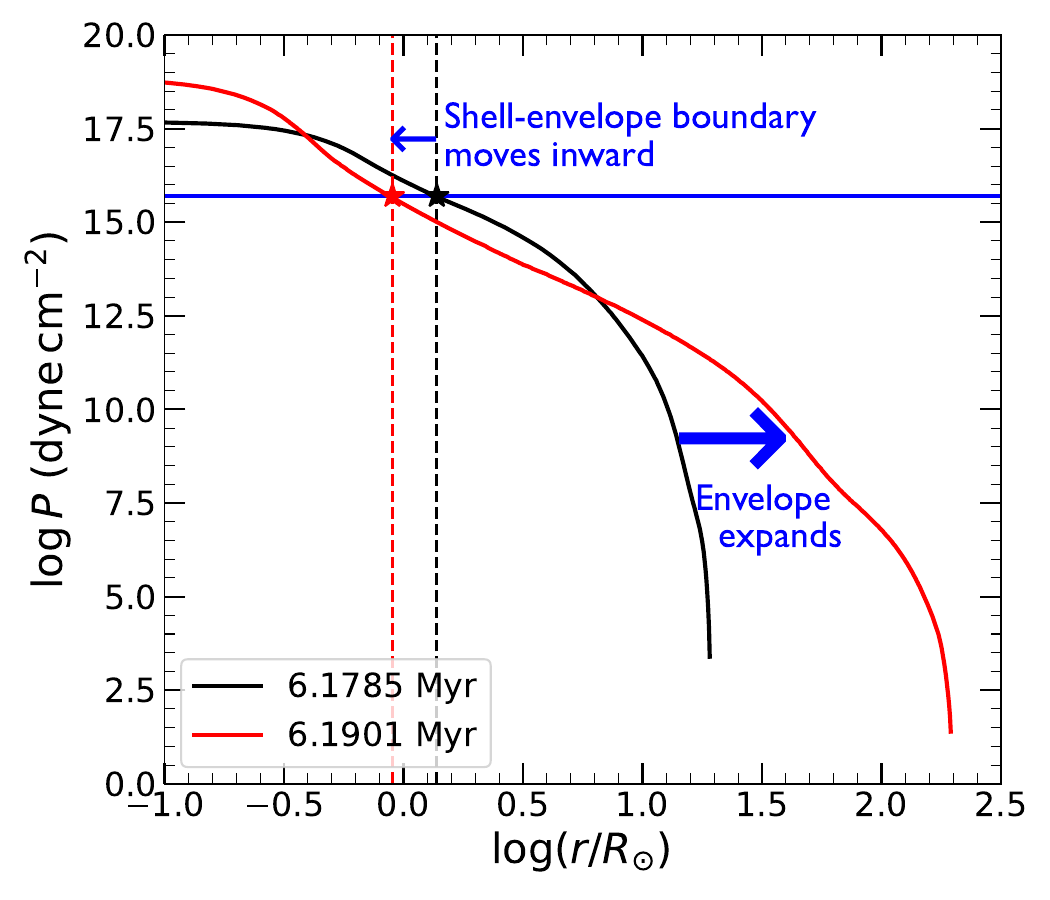}
\includegraphics[scale=0.45]{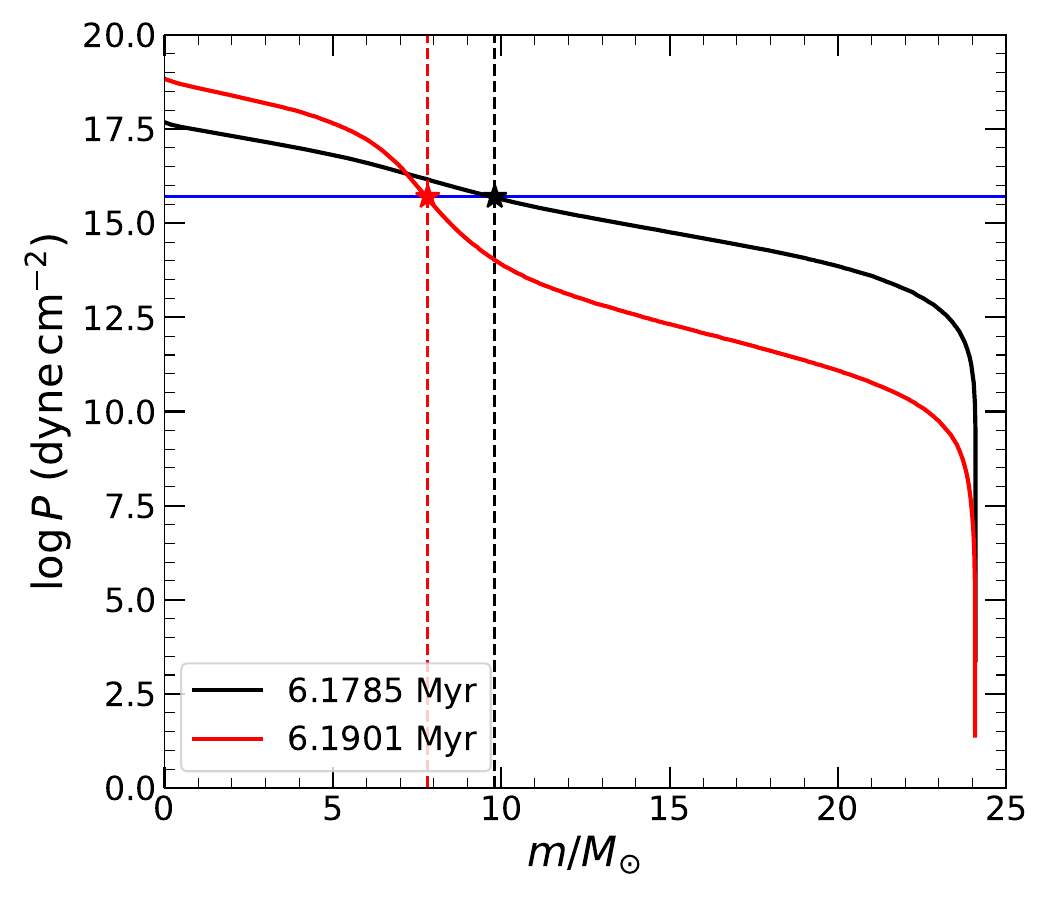}
\includegraphics[scale=0.45]{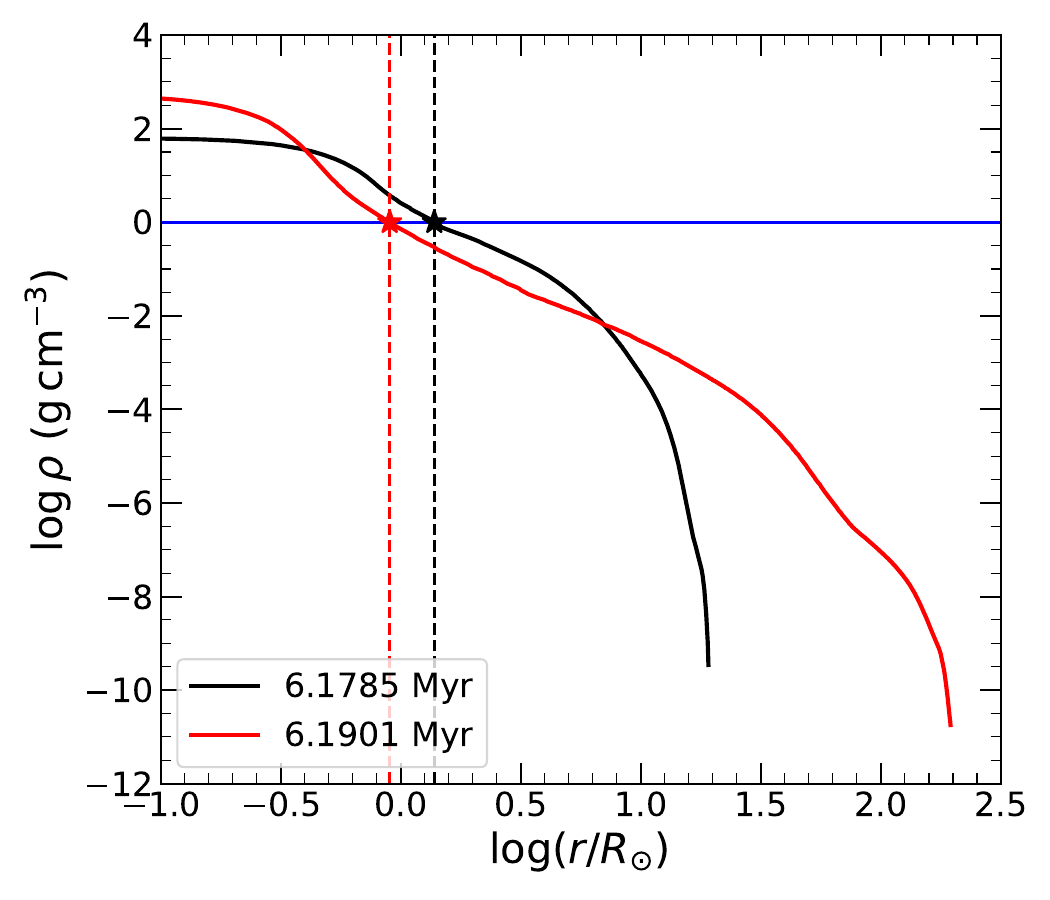}
\includegraphics[scale=0.45]{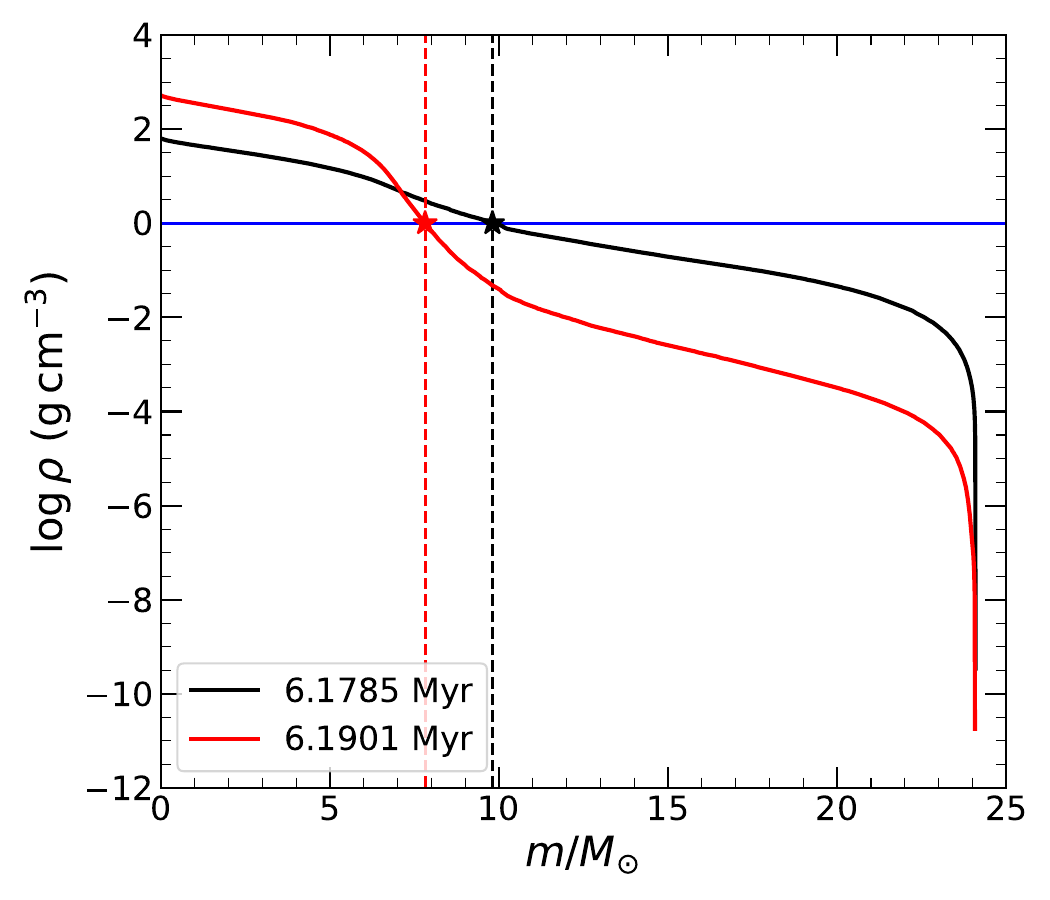}

\caption{Pressure (\textit{top}) and density (\textit{bottom}) profiles at two timesteps for a $25\,M_{\odot}$, $Z=0.02$ star during envelope expansion, shown in radius (\textit{left}) and mass (\textit{right}) coordinates. Dashed lines mark the outer edge of the H-burning shell, the inner envelope boundary. Star symbols indicate nearly constant pressure and density at this boundary (blue horizontal lines). As the shell-envelope boundary moves inward, the inner envelope's pressure gradient ($dP/dm$) steepens to maintain hydrostatic equilibrium, reducing pressure and density throughout the envelope and driving its expansion.
}
\label{fig:P_profile}
\end{figure*}

Based on the nearly constant pressure at the envelope’s inner boundary, the refined mirror principle can be understood from the equation of hydrostatic equilibrium,

\begin{equation}
\frac{dP}{dm} = -\frac{Gm}{4\pi r^4},
\end{equation}

\noindent where $P$ is the local pressure, $m$ is the enclosed mass, $r$ is the radial coordinate, and $G$ is the gravitational constant. Fig.~\ref{fig:P_profile} illustrates the pressure and density profiles at two time steps during the envelope expansion of a $25\,M_{\odot}$, $Z = 0.02$ model, plotted in both radius and mass coordinates. Between these times, the shell-envelope boundary remains at nearly constant pressure while moving inward in both radius and mass. As the central region of the star contracts, $R_{\rm shell}^4$ decreases much more rapidly (by $\sim$80\%) than the enclosed mass at the same location, $M_{\rm shell}$ (by $\sim$20\%). Therefore, according to Eq.~(7), hydrostatic equilibrium requires a steeper pressure gradient, $|dP/dm|$. With the pressure fixed at the envelope’s inner boundary, the pressure within the overlying envelope must decrease to achieve this greater gradient. The envelope density consequently decreases according to the equation of state, leading to its expansion. Hence, the inward motion of the shell-envelope boundary naturally drives envelope expansion, providing a physical basis for the refined mirror principle.

Unlike previous interpretations based on energy or virial arguments, the refined mirror principle arises directly from hydrostatic equilibrium under the condition of nearly fixed pressure at the shell-envelope boundary. Paper~II \citep{paper2} further validates this principle by solving the steady-state hydrostatic envelope structure.

%%%%%%%%%%%%%%%%
%%%%%%%%%%%%%%%%

\section{Effects of Envelope Properties}\label{sec:envelope}

We have shown that the core and shell properties largely govern whether the envelope expands into the RSG phase. Nevertheless, given that previous studies often attribute RSG formation to intrinsic envelope properties, we now examine the roles of envelope characteristics, focusing on opacity and internal structure.

\subsection{The Role of Opacity}
Earlier studies based on thermal imbalance suggested that envelope expansion is driven by heat absorption, with higher core luminosities and greater envelope opacity favoring the transition to RGs/RSGs \citep{Renzini1984, Renzini1992, Renzini1994, Ritossa1996, Renzini2023}. This view was challenged by \citet{Iben1993}, who argued that heat absorption in the envelope is a consequence of convective envelope development, not the cause of expansion.

\begin{figure*}[tbh]
\centering
\includegraphics[scale=0.35]{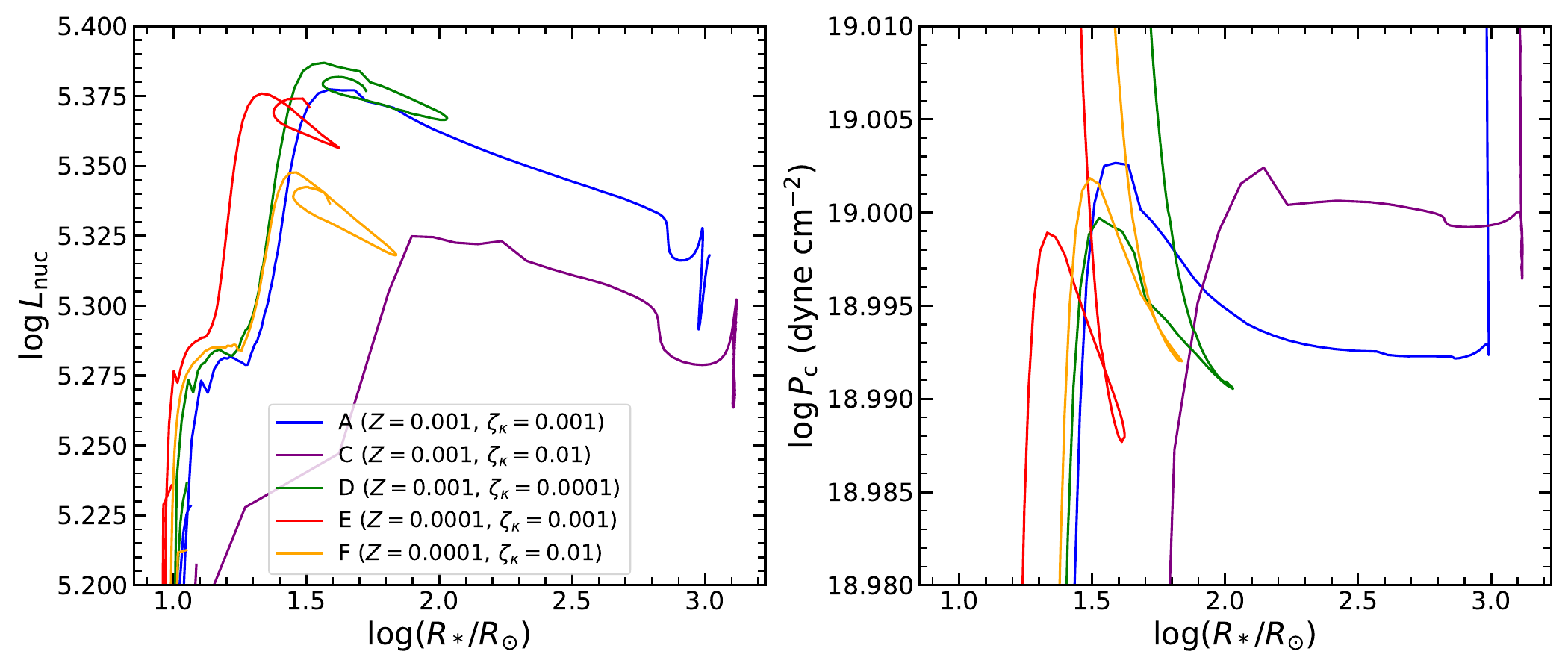}
\caption{Evolution of the nuclear energy generation rate ($L_{\rm nuc}$) and central pressure ($P_{\rm c}$) as functions of stellar radius ($R_*$) for models with varying metallicities ($Z$) and opacity parameters ($\zeta_{\kappa}$). Arrows indicate the direction of evolution. Differences in $L_{\rm nuc}$ and the location of the $P_{\rm c}$ peak illustrate how opacity and metallicity influence envelope expansion toward the RSG phase.
}
\label{fig:z-zbase}
\end{figure*}

In Section~\ref{sec:criterion}, we showed that expansion toward RSGs cannot be explained solely by increased energy supply. We now examine the combined roles of core luminosity and envelope opacity by constructing models at two metallicities ($Z=0.001$ and $0.0001$) with three values of the opacity parameter $\zeta_{\kappa}$ ($0.01$, $0.001$, and $0.0001$; see Section~\ref{sec:methods}). The model with $Z=0.001$ and $\zeta_{\kappa}=0.001$ corresponds to Model~A discussed previously. Fig.~\ref{fig:z-zbase} shows the evolution of $L_{\rm nuc}$ and $P_{\rm c}$ as functions of $\log R_*$ for Models~A, C, D, E, and F.

Comparing Models A and F, we see that despite Model F having higher opacity, it does not expand into an RSG, whereas Model A does. This indicates that higher opacity alone does not guarantee RSG formation. Under the heat absorption scenario, one might argue that Model F fails to expand because of its lower $L_{\rm nuc}$ and, consequently, an insufficient energy supply. However, a comparison between Models F and C challenges this interpretation: although both share the same opacity-scaling parameter $\zeta_{\kappa}$, they differ in metallicity $Z$, leading to different $L_{\rm nuc}$. Model C, with a lower $L_{\rm nuc}$, should in principle be even less likely to expand than Model F, yet it evolves into an RSG—contradicting the energy-driven expansion hypothesis. Thus, the heat-absorption scenario cannot consistently explain RSG formation.

While higher opacity does not directly drive envelope expansion through energy trapping, it modulates the stellar structure and thereby indirectly influencing the pathway toward the RSG phase. Fig.~\ref{fig:z-zbase} shows that higher opacity shifts the peak of $P_{\rm c}$ to larger $R_*$. For example, at the same $Z$, Model~A (lower $\zeta_{\kappa}$) reaches its pressure peak at $R_* \sim 30$–$40\,R_{\odot}$, whereas Model~C (higher $\zeta_{\kappa}$) peaks at $R_* > 100\,R_{\odot}$. Hence, at a comparable core-evolution stage, a model with higher opacity tends to possess a larger envelope radius. The envelope radius at a given core stage is crucial in determining the supergiant evolutionary pathway, a subject explored in detail in \citet{Ou2025}.

%%%%%%%%%%%%%%%%
%%%%%%%%%%%%%%%%

\subsection{Structural Transition into the RG/RSG Phase}

\begin{figure*}[tbh]
\centering
\includegraphics[width=\textwidth]{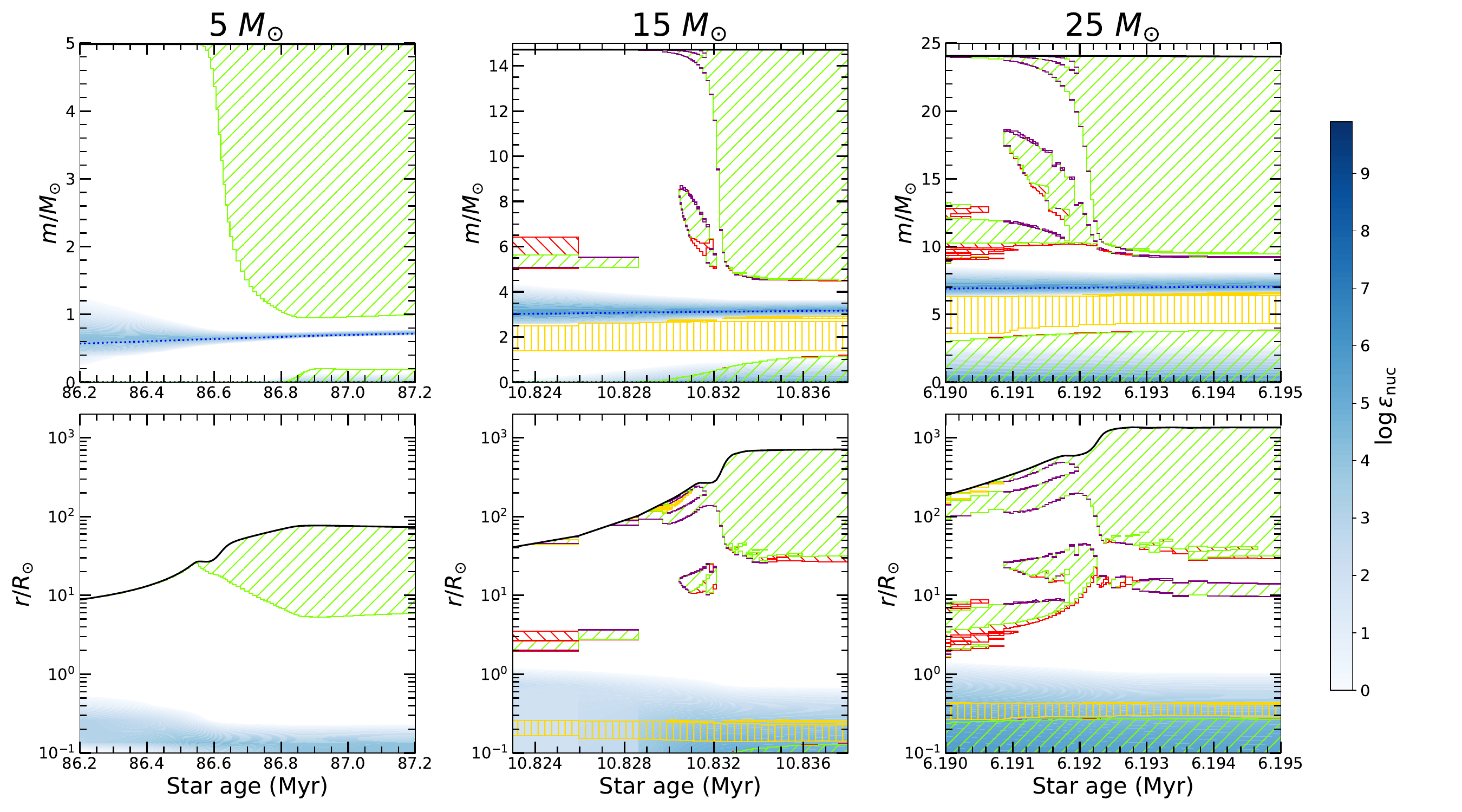}
\caption{Kippenhahn diagrams for $5$, $15$, and $25\,M_{\odot}$ stars at $Z=0.02$. Top panels use mass coordinates, bottom panels radius coordinates. Striped regions indicate mixing: green = convection, red = semi-convection, yellow = thermohaline, purple = overshooting; blue background shows $\log \epsilon_{\rm nuc}$ (erg s$^{-1}$ g$^{-1}$). Diagrams are zoomed on the RG/RSG transition phase, highlighting the development of an extended convective envelope. A brief pause in radial expansion occurs just before convection onset at stellar ages of $\sim 86.55$, $10.831$, and $6.192$ Myr for the $5$, $15$, and $25\,M_{\odot}$ models, respectively.
}
\label{fig:kipp}
\end{figure*}

While core and shell properties primarily determine whether a star evolves into a BSG or RSG, the envelope structure becomes crucial in the final approach to the RSG phase. Kippenhahn diagrams in Fig.~\ref{fig:kipp} reveal a brief pause in radial expansion for each model (at stellar ages of $\sim$ 86.55, 10.831, and 6.192 Myr for the 5, 15, and $25\,M_{\odot}$ models, respectively). Once the star resume expansion after this brief pause, its surface convective zone rapidly deepens inward to the base of the envelope.

\begin{figure*}[tbh]
\centering
\includegraphics[scale=0.45]{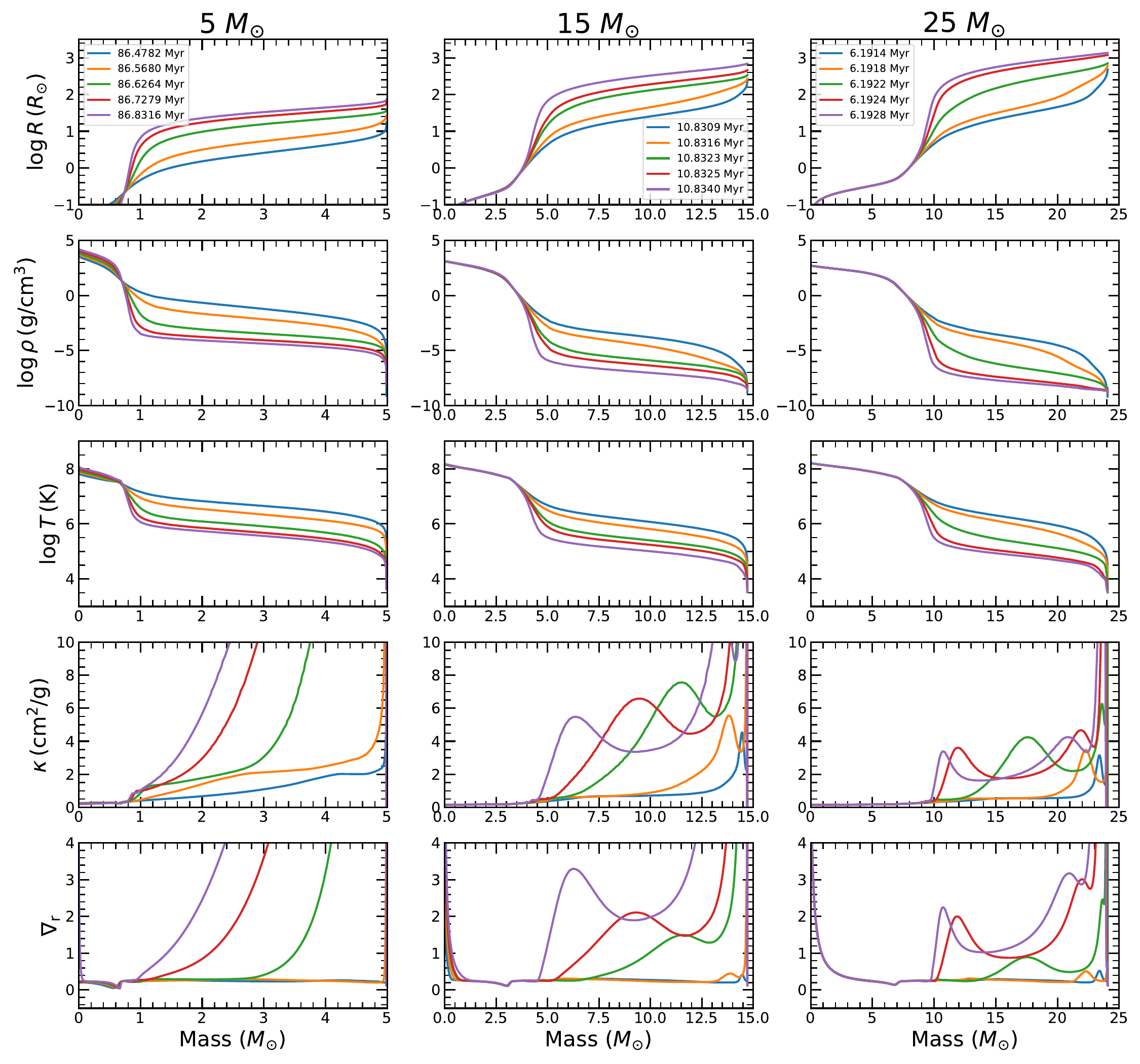}
\caption{Profiles of radius, density, temperature, opacity, and radiative temperature gradient ($\nabla_{\rm r}$) in mass coordinates during the RG/RSG transition for $5$, $15$, and $25\,M_{\odot}$ stars at $Z=0.02$. During this phase, envelope mass redistributes outward, cooling most of the stellar mass, increasing opacity, steepening $\nabla_{\rm r}$, and triggering the formation of an extended convective zone.
}
\label{fig:mass5profiles}
\end{figure*}

\begin{figure*}[tbh]
\centering
\includegraphics[scale=0.35]{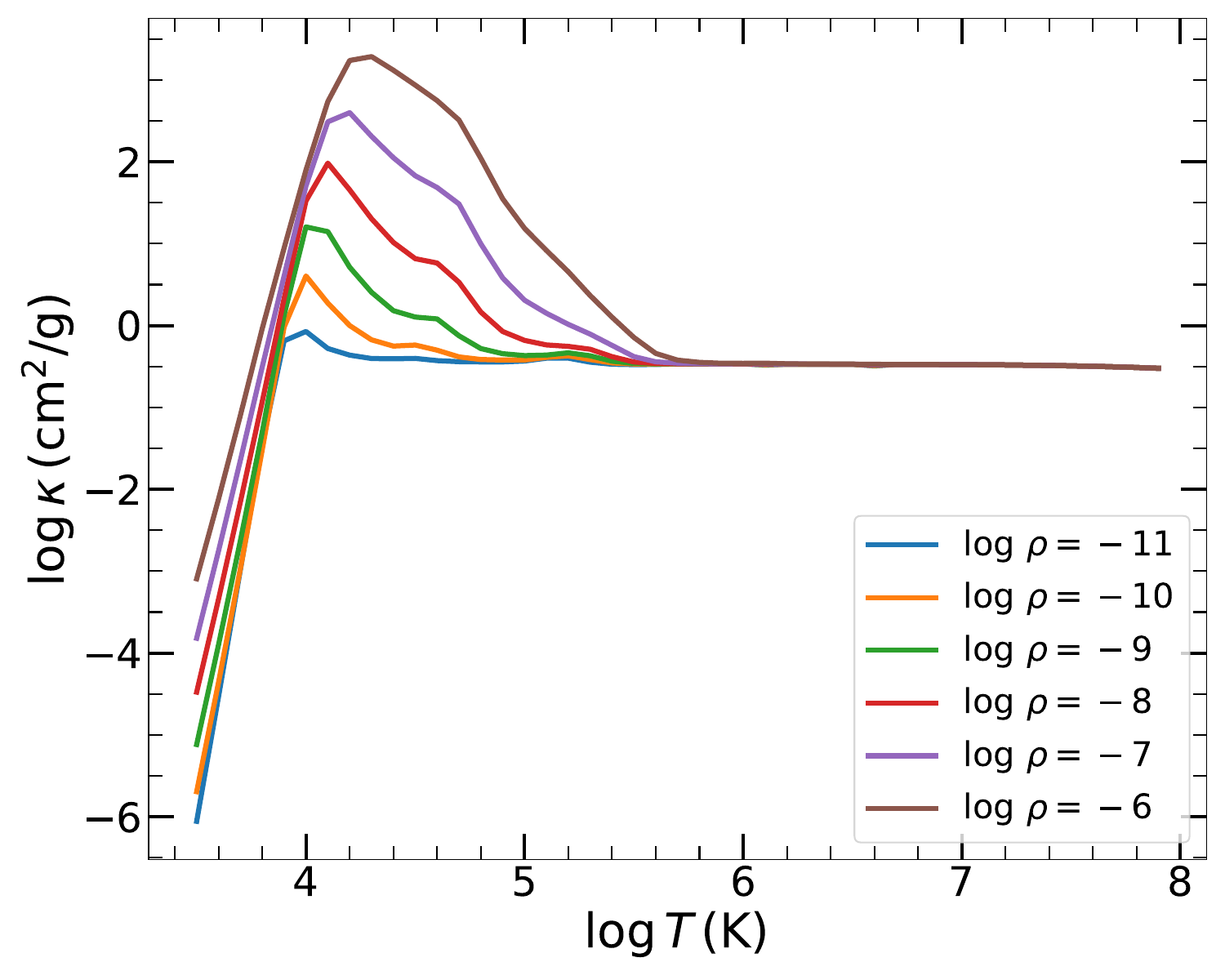}
\caption{Opacity as a function of temperature and density from the "gs98" opacity table in MESA. As stellar material cools from high temperatures to a few times $10^5$ K, the opacity rises steeply with decreasing temperature, following Kramers’ law.}
\label{fig:kappa_profile}
\end{figure*}

During this brief pause of radial expansion, the profiles of radius, density, temperature, opacity, and radiative gradient (Fig.~\ref{fig:mass5profiles}) exhibit significant changes. Before the transition, only a small fraction of the envelope mass near the surface extends to large radii, while most of the envelope remains compact. During the transition, however, the bulk of the mass above the H-burning shell redistributes outward, forming a low-density, extended envelope. This mass redistribution cools most of the envelope to $\lesssim$ a few $10^5$ K, sharply increasing opacity (Fig.~\ref{fig:kappa_profile}). The higher opacity raises the radiative gradient, thereby triggering an extended convective zone that penetrates deep into the base of the envelope.

\begin{figure*}[tbh]
\centering
\includegraphics[scale = 0.5]{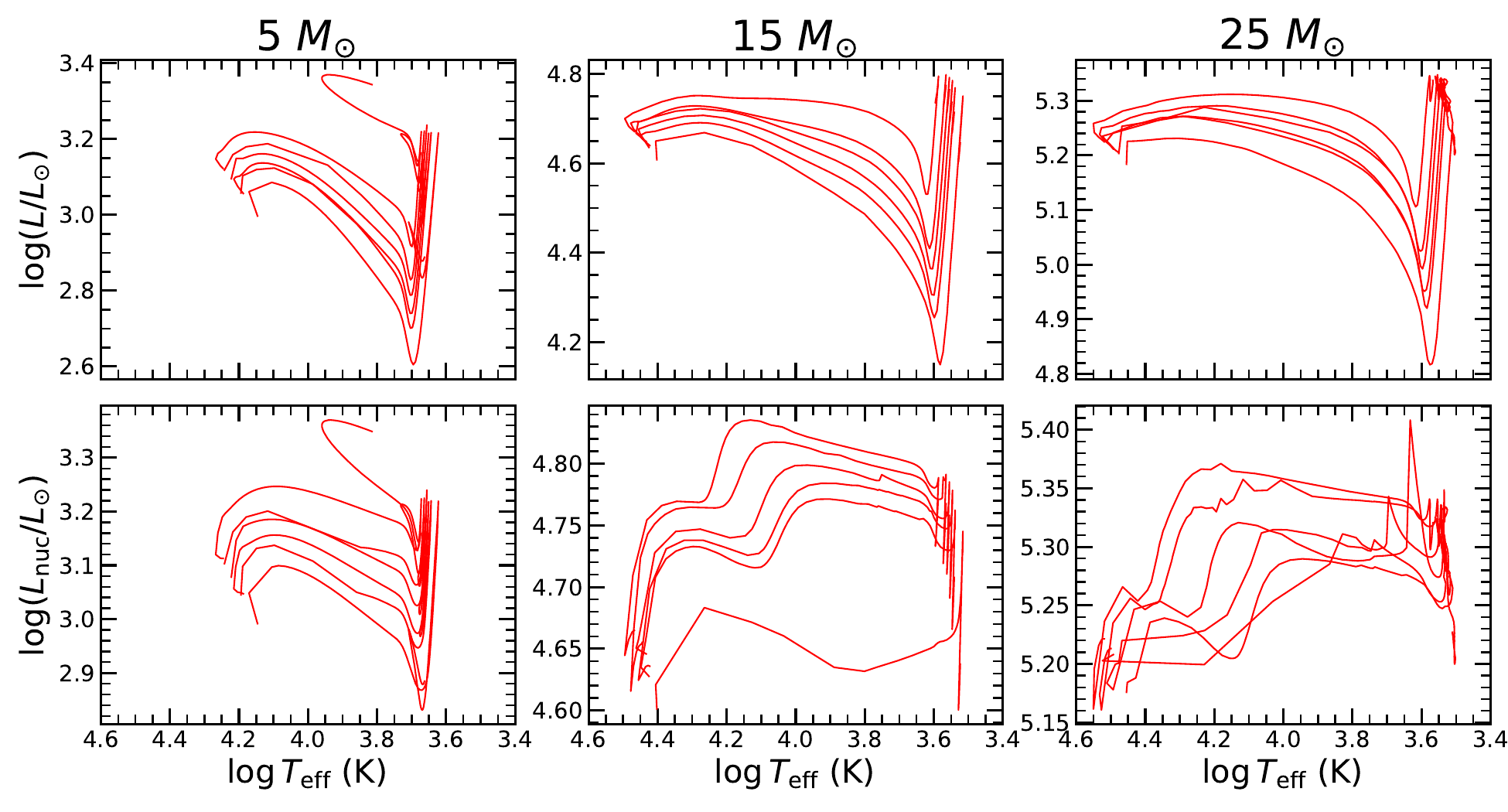}
\caption{Evolution of surface luminosity ($L_s$) and nuclear energy generation rate ($L_{\rm nuc}$) as functions of effective temperature for $5$, $15$, and $25\,M_{\odot}$ stars with metallicities $Z = 0.002$–0.02. Lower-metallicity models generally exhibit higher $L_s$ and $L_{\rm nuc}$. A sharp decline in both quantities occurs as $T_{\rm eff}$ drops to $\sim 4,000$–5,000 K, marking the transition into the RG/RSG phase. Evolution proceeds from higher to lower $T_{\rm eff}$.
}
\label{fig:Lnuc-5-15-25}
\end{figure*}

This structural transition explains the behavior of $L_s$ and $L_{\rm nuc}$ just before the star enters the RSG regime, as shown in Fig.~\ref{fig:Lnuc-5-15-25}: a sharp decline followed by a rapid recovery. The luminosity drop occurs not only because energy is consumed to drive envelope expansion, but also because the sudden increase in opacity within the envelope during the transition further inhibits energy transport. Once the extended convective zone is established, efficient energy transport is restored, allowing the luminosity to recover.

In summary, the radial expansion triggers a structural phase transition in the envelope just before the star reaches RG/RSG dimensions. This transition establishes a convective envelope and stabilizes the star at RG/RSG sizes, demonstrating that the RG/RSG configuration represents a distinct structural phase from the BG/BSG stage. Paper~II \citep{paper2} reproduces this transition using steady-state envelope models and elaborates on its physical significance.

%%%%%%%%%%%%%%%%
%%%%%%%%%%%%%%%%
%%%%%%%%%%%%%%%%
%%%%%%%%%%%%%%%%
%%%%%%%%%%%%%%%%
%%%%%%%%%%%%%%%%
%%%%%%%%%%%%%%%%
%%%%%%%%%%%%%%%%

\section{Conclusions}\label{sec:conclusion}

In this study, we revisit the long-standing question: Why do stars evolve into RGs/RSGs? We identify the primary mechanism governing envelope expansion in post-main-sequence stars as a refined form of the ``mirror principle.'' Unlike the traditional view, in which the burning shell mirrors core contraction with envelope expansion, we find that the envelope consistently expands or contracts in the opposite direction of its inner boundary, defined by the outer edge of the H-burning shell, manifesting as a strong anti-correlation between $R_*$ and $R_{\rm shell}$.

This refined mirror principle arises naturally from hydrostatic equilibrium and mass conservation, rather than from energy absorption in the envelope. The shell-envelope boundary maintains nearly constant temperature and pressure. As the boundary moves inward, gravitational acceleration increases, requiring a steeper pressure gradient. With the pressure at the base fixed, the envelope responds by reducing density and temperature through expansion. In Paper~II \citep{paper2}, we validate this mechanism by solving steady-state stellar structure equations. How $L_{\rm nuc}$ and other stellar parameters control $R_{\rm shell}$ will be addressed in future studies.

We emphasize that thermal imbalance and energy absorption, often invoked to explain supergiant formation, are consequences rather than causes of expansion. The envelope's extent is determined by the configuration required to maintain hydrostatic equilibrium, not by the amount of energy supplied.

Applying the refined mirror principle, we propose a general roadmap for post-main-sequence evolution toward the RG/RSG phase (Fig.~\ref{fig:flow}), comprising two pathways:

\paragraph{(1) Expansion during core contraction.} After core H exhaustion, the He core contracts, establishing an H-burning shell. As the shell moves inward with the contracting core, the envelope expands outward. In many cases, this alone is sufficient for the star to reach RG/RSG dimensions.
\paragraph{(2) Expansion after core contraction.} If the star remains a BG or BSG after He-core contraction, continued envelope expansion depends on nuclear burning. As $L_{\rm nuc}$ declines, the shell moves inward and the envelope expands. If $L_{\rm nuc}$ rises, the shell reverses, and the envelope contracts, stabilizing as a BG/BSG. Only a sustained decline in $L_{\rm nuc}$ allows expansion to continue into the RG/RSG phase.

Thus, RG/RSG formation can occur either directly during core contraction or afterward, regulated by nuclear burning. In both cases, the expansion is governed by the refined mirror principle. The pathway a star follows depends on the correspondence between the evolutionary stages of its envelope and core, which is examined in detail in a another paper \citet{Ou2025}.

We also identify a structural change during the final stage of envelope expansion that leads to the phase transition into the RG/RSG configuration, whose internal structure is distinct from that of the BG/BSG stage. During this transition, most of the envelope mass redistributes outward and cools, increasing the opacity and steepening the radiative temperature gradient. This triggers the formation of a deep, extended convective zone and produces a sharp dip in surface luminosity, marking the completion of the envelope’s expansion.

This study focuses on expansion prior to core-He exhaustion; later-stage expansion may influence phenomena such as Case C mass transfer in binaries \citep{Lauterborn1970, Podsiadlowski1992}, warranting future investigation. 

In summary, we present a unified physical framework based on the refined mirror principle that explains why post-main-sequence stars evolve into the RG/RSG phase. Paper~II \citep{paper2} further validates the refined mirror principle and the structural transition leading to the RG/RSG phase using steady-state envelope models.

\begin{acknowledgements} 
This research is supported by the National Science and Technology Council, Taiwan, under grant No. NSTC 113-2112-M-001-028-, 114-2811-M-001-094-, 114-2112-M-001-012-, and the Academia Sinica, Taiwan, under a career development award under grant No. AS-CDA-111-M04. KC acknowledges the support of the Alexander von Humboldt Foundation and Heidelberg Institute for Theoretical Studies. This research was supported in part by grant NSF PHY-2309135 to the Kavli Institute for Theoretical Physics (KITP) and grant NSF PHY-2210452 to the Aspen Center for Physics. Our simulations were performed at the National Energy Research Scientific Computing Center (NERSC), a U.S. Department of Energy Office of Science User Facility operated under Contract No. DE-AC02-05CH11231, and on the TIARA Cluster at the Academia Sinica Institute of Astronomy and Astrophysics (ASIAA).
\end{acknowledgements}

%\software{MESA \citep{Paxton2011,Paxton2013,Paxton2015,Paxton2018,Paxton2019,Jermyn2023}

\bibliography{RSG}{}
\bibliographystyle{aa}

%\end{CJK*}
\end{document}